\documentclass[preprint2]{aastex61}
\usepackage{amsmath}
\usepackage{txfonts}
\usepackage{graphicx}
\usepackage[pdf]{pstricks}
\usepackage{xcolor}
\usepackage{dsfont}
\usepackage{natbib}
\usepackage{float}
\usepackage{nicefrac}
\usepackage{afterpage}
\usepackage{mathrsfs}

\DeclareMathAlphabet{\mathsfit}{\encodingdefault}{\sfdefault}{m}{sl}
\SetMathAlphabet{\mathsfit}{bold}{\encodingdefault}{\sfdefault}{bx}{sl}

\newcommand{\mytensor}[1]{\boldsymbol{\mathsfit{#1}}}

\begin{document}
\title{Structure of the solar photosphere studied from the radiation hydrodynamics code \texttt{ANTARES}}
\shorttitle{Structure of the solar photosphere studied from the RHD model ANTARES}
\shortauthors{P.~Leitner et al.}

\author{P.~Leitner}
\affiliation{University of Graz, Universit\"atsplatz 5, A-8010~Graz, Austria}
\email{peter.leitner@alumni.uni-graz.at}

\author{B.~Lemmerer}
\affiliation{University of Graz, Universit\"atsplatz 5, A-8010~Graz, Austria}

\author{A.~Hanslmeier}
\affiliation{University of Graz, Universit\"atsplatz 5, A-8010~Graz, Austria}

\author{T.~Zaqarashvili}
\affiliation{University of Graz, Universit\"atsplatz 5, A-8010~Graz, Austria}
\affiliation{Space Research Institute, Austrian Academy of Sciences, Schmiedlstrasse 6, A-8042~Graz, Austria}
\affiliation{Abastumani Astrophysical Observatory at Ilia State University, 3/5 Cholokashvili avenue, 0162 Tbilisi, Georgia}

\author{A.~Veronig}
\affiliation{University of Graz, Universit\"atsplatz 5, A-8010~Graz, Austria}

\author{H.~Grimm-Strele}
\affiliation{Faculty of Mathematics, University of Vienna, Nordbergstrasse~15, A-1090~Wien, Austria}
\affiliation{Max-Planck Institute for Astrophysics, Karl-Schwarzschild-Strasse~1, 85741~Garching, Germany}

\author{H.J.~Muthsam}
\affiliation{Faculty of Mathematics, University of Vienna, Nordbergstrasse~15, A-1090~Wien, Austria}

\begin{abstract}
The \texttt{ANTARES} radiation hydrodynamics code is capable of simulating the solar granulation in detail unequaled by direct observation. We introduce a state-of-the-art numerical tool to the solar physics community and demonstrate its applicability to model the solar granulation. The code is based on the weighted essentially non-oscillatory finite volume method and by its implementation of local mesh refinement is also capable of simulating turbulent fluids. While the \texttt{ANTARES} code already provides promising insights into small-scale dynamical processes occurring in the quiet-Sun photosphere, it will soon be capable of modeling the latter in the scope of radiation magnetohydrodynamics. In this first preliminary study we focus on the vertical photospheric stratification by examining a 3-D model photosphere with an evolution time much larger than the dynamical timescales of the solar granulation and of particular large horizontal extent corresponding to $25\arcsec \times 25\arcsec$ on the solar surface to smooth out horizontal spatial inhomogeneities separately for up- and downflows. The highly resolved Cartesian grid thereby covers $\sim 4~\mathrm{Mm}$ of the upper convection zone and the adjacent photosphere. Correlation analysis, both local and two-point, provides a suitable means to probe the photospheric structure and thereby to identify several layers of characteristic dynamics: The thermal convection zone is found to reach some ten kilometers above the solar surface, while convectively overshooting gas penetrates even higher into the low photosphere. An $\approx 145\ \mathrm{km}$ wide transition layer separates the convective from the oscillatory layers in the higher photosphere.
\end{abstract}


\section{Introduction}
The structure and dynamics of the solar photosphere is crucially determined by the mass and energy transport processes taking place across the solar surface. For the study of the solar convection and its phenomenological manifestation on the visible surface of the Sun, the solar granulation, numerical simulations not only complement observational data but also serve as a means of their own by providing complete and almost continuous information in 3-D of the physical state and the dynamics which otherwise often has
to be drawn indirectly from observations. The physics of the layers surrounding the solar surface is rather involved; below the surface the opacity is sufficiently large so that the local adiabatic gradient is exceeded by the temperature gradient needed for radiative-diffusive energy transport, turning the fluid convectively unstable. This process is primarily described by mixing length theory \citep[e.g.][]{Cox1968,Spruit1974} that proved to be successful at determining the average energy transport \citep{Cattaneo1991}, while neglecting e.g. the dynamical modifications to the hydrostatic equilibrium near the surface \citep{Macgregor1991}.

\begin{figure*}
\begin{center}
\includegraphics[width=0.43\textwidth]{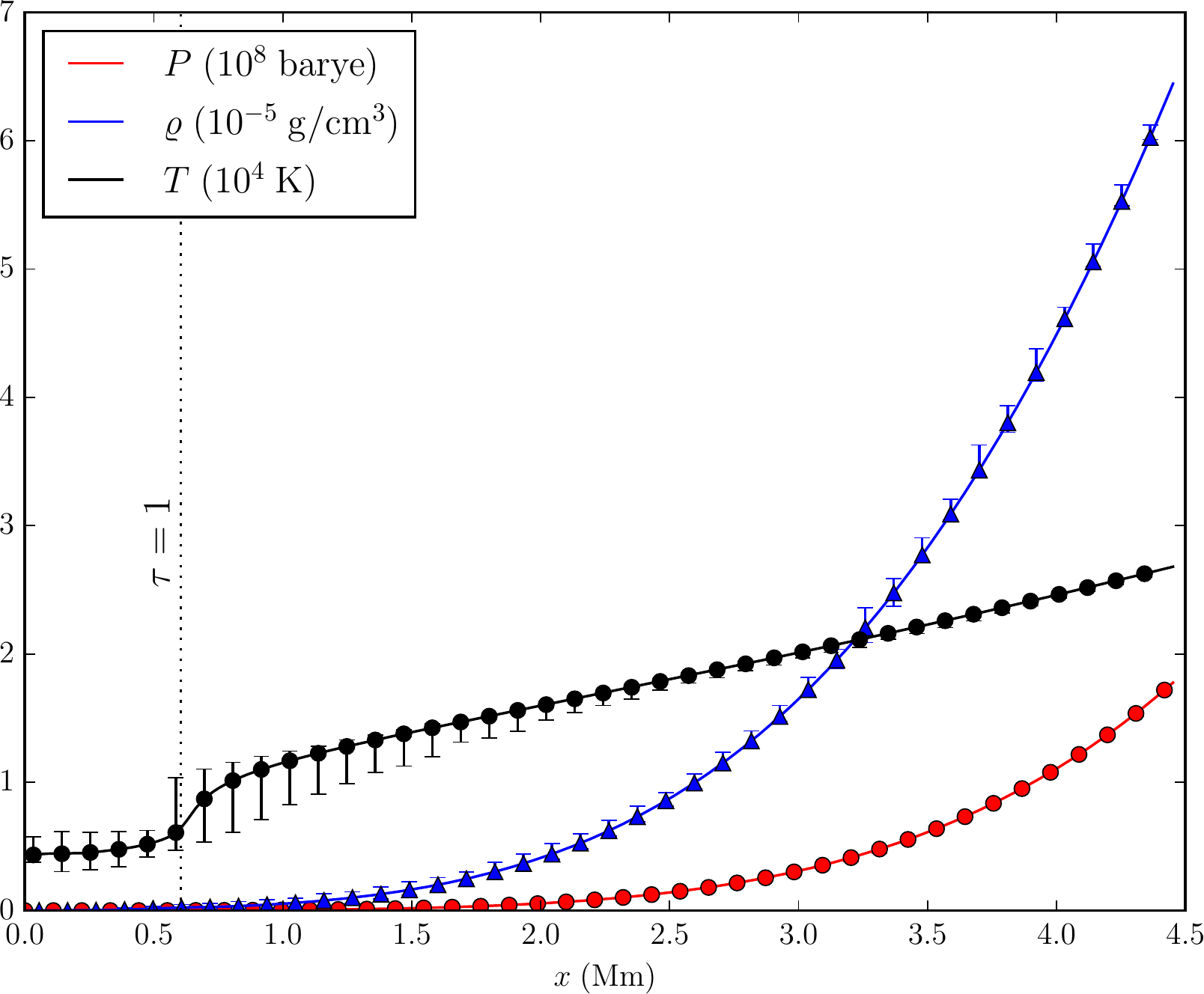}\hspace{0.3cm}
\includegraphics[width=0.45\textwidth]{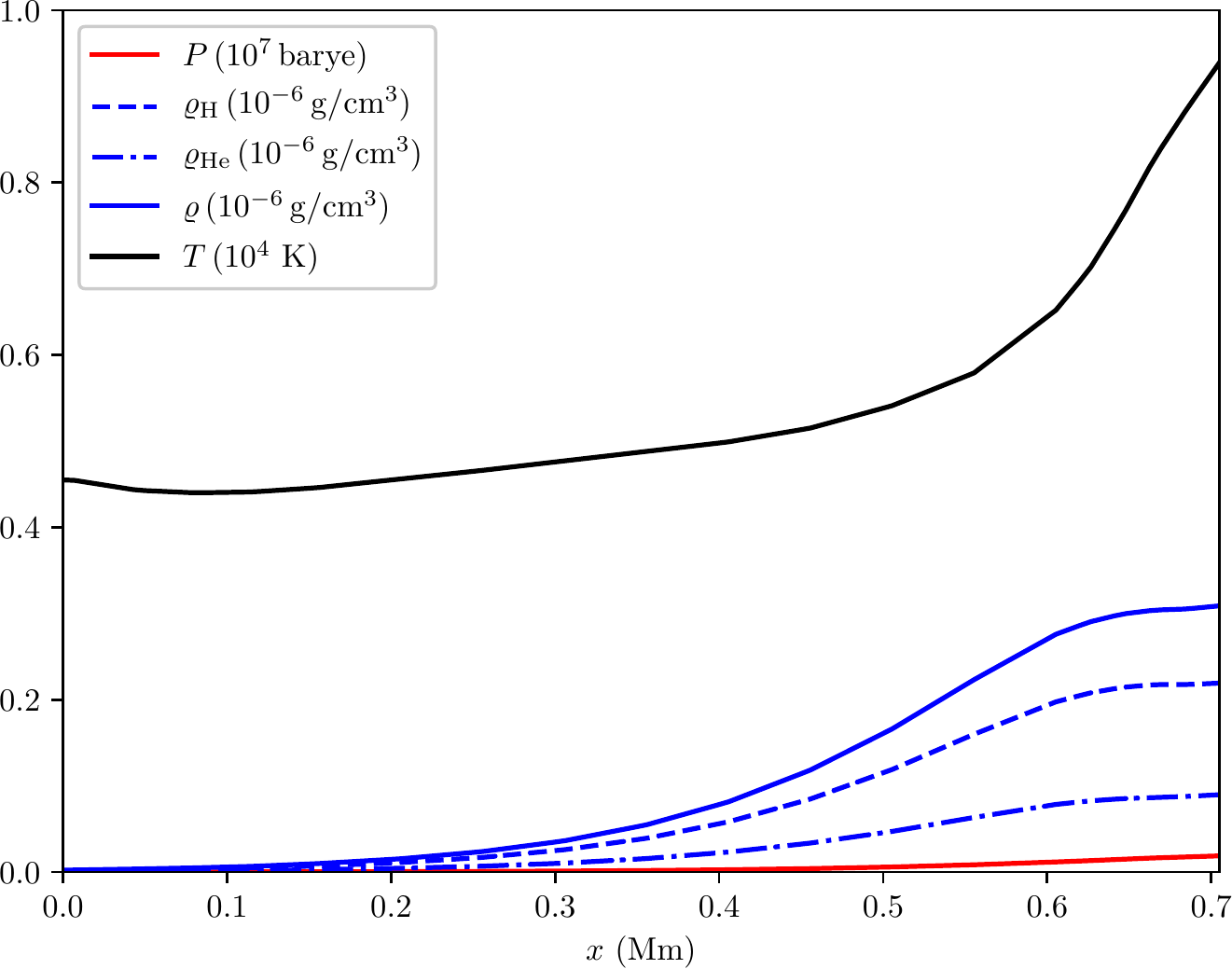}
\end{center}
\vspace{-4pt}
\caption{\textit{Left}: Vertical profile of horizontally averaged pressure, density, and temperature for a given model time step as a function of geometric depth $x$ as obtained from the \texttt{ANTARES} model based on the \texttt{ATLAS 9} package \citep{Kurucz1970}. Error-bars indicate the quantities' variation at a given geometric depth. \textit{Right}: Comparison to data from the energy-balance model atmosphere No.~C of the
quiet Sun of \citet{Fontenla1993}. The steep temperature gradient at the surface that is visible in both model atmospheres is due to the large temperature sensitivity of the dominant $\mathrm{H}^-$ opacity \citep[e.g][]{Stein1998}}
\label{fig:vertical_runs}
\end{figure*}

At the solar surface and above, along with a rapid decrease of the opacity, radiation becomes the primary energy transport mechanism, although a continuing mass flow overshooting into convectively stable regions still characterizes the lower photosphere. Ionization and molecular dissociation processes taking place in the near surface layers considerably affect the internal energy and the equation of state. The extent to which numerical models reflect reality has improved significantly with increasing numerical accuracy and level of physical detail since the 1970s, one such enhancement being the consideration of non-gray radiative transport.

The photosphere and the subjacent thin $4~\mbox{Mm}$ wide layer of the upper convection zone that are covered in the simulation studied here constitute a region of almost negligible radial extent, accounting for no more than 0.64\% of the solar radius. Yet it is these layers that feature remarkably rich dynamical processes driven by steep vertical gradients of temperature, pressure, and density, the latter falling off by 5 and 4 orders of magnitude, respectively, see Fig.~\ref{fig:vertical_runs}.

We present the state-of-the-art radiation hydrodynamics (RHD) code \texttt{ANTARES} (A~Numerical Tool for Astrophysical RESearch), applied to the study of the near surface convection and the photosphere of the Sun. In spite of the broad range of applicability, ranging from the modeling of photospheric turbulence \citep{Muthsam2007} to Cepheid pulsation \citep{Muthsam2011}, the code has not received much attention in the solar physics community so far. While, for the time being, the code is restricted to the modeling of the quiet Sun, an MHD upgrade is foreseen in the near future that will allow us to study the modifications to the dynamics and to the energy transport due to the photospheric magnetic field and to extend its applicability to magnetoactive regions at the solar surface. Further recent developments of the code are summarized in \citet{Blies2015} and include the consideration of two-component flows \citep{Zaussinger2010}, a parallel multigrid solver for the 2-D non-linear Helmholtz equation \citep{Happenhofer2013}, and a generalization to solve the Navier-Stokes equations on curvilinear grids \citep{GrimmStrele2014}. While the over many years of development fully matured \texttt{ANTARES} code is characterized by its elaborate numerical schemes and its stability, it is by far not the only simulation project of similar scope of applicability. Of the many other codes we can only exemplarily mention a few such as the 3-D MHD code of \citet{Nordlund1997} that lead the way in many respects and by which it was first possible to model the solar atmosphere from the photosphere up to the corona \citep{Gudiksen2002}. Of late the code CO5BOLD, developed at the Kiepenhauer Institut
\citep[e.g.][]{Freytag2010} has gained a lot of attention in connection with the observation of rapidly rotating magnetic structures, so-called magnetic tornadoes, in chromospheric simulation data that serve as an energy channel into the corona \citep{Wedemeyer2012}. It not so long ago obtained an optional generalization from RHD to MHD \citep{Schaffenberger2005} and, in conjunction with the related spectral analysis package
Linfor3D \citep{Gallagher2016} is also capable of computing the emergent spectrum, allowing for a more direct comparison to observations. The interaction of magnetic fields with radiative convection and further magnetic activity in the photosphere, such as the formation and dynamics of magnetic flux tubes are intensely studied with the MURaM code \citep[e.g.][]{Voegler2005}, that is a collaborative project of the Max-Planck Institute for Solar System Research and the University of Chicago. Finally, the young computational heliophysics 2-fluid code JOANNA that is developed at the Maria Curie-Sk\l odowska University in Lublin, Poland already revealed promising results concerning the simulation of spicules, showing that triggering pulses steepen into upward propagating shocks whereas the chromospheric cold and dense plasma lags behind such shocks during its rise into the corona with a mean speed of $20\ldots 25~\mbox{km}/\mbox{s}$ \citep{Kuzma2017}. This is neither a ranking list nor should it be considered to be anything but incomplete, its sole purpose here being to give an impression of the industriousness of this particular research area and of the vast efforts that are put into the examination of the complex energy transport mechanisms in  the solar atmosphere with i.a. the intention to finally fully understand the heating of the solar corona.

The high resolution of the \texttt{ANTARES} model photospheres allows further detailed studies of smaller-scale dynamical phenomena such as quiet-Sun jets that have been reported from the IMaX instrument on board of the SUNRISE stratospheric balloon telescope \citep{MartinezPillet2011,Borrero2010}, jet-like vortex tubes that have been observed with the New Solar Telescope (NST) by \citet{Yurchyshyn2011}, or fine-scaled MHD phenomena occurring in MHD simulations \citep{Kitiashvili2014} that cause intense dynamic interactions between the surface and the chromosphere and which may be responsible for acoustic wave excitation and quasi-periodic flow eruptions, see e.g. \citet{Kitiashvili2013} and references listed therein. The present RHD model has already revealed rotating plasma jets \citep{Lemmerer2016}, which seem to be triggered by turbulent convection. Our current investigations \citep{Lemmerer2016} suggest that horizontal flows around rotating jets may trigger MHD kink waves \citep[e.g.][]{Zaqarashvili2007} or torsional Alfv\'{e}n waves \citep{Fedun2011,Shelyag2013,Zaqarashvili2013} that may propagate in magnetic flux tubes that are anchored in the photosphere and thereby transport photospheric energy into the chromosphere. First approximations showed that the observed kinetic energy flux may excite waves carrying an energy flux of $\sim 7 \times10^7~\mbox{erg}\,\mbox{cm}^{-2}\,\mbox{s}^{-1}$, supposing only a 1\% energy transfer to the waves. This wave energy flux is about one order higher than necessary to compensate for energy losses from the quiet-Sun chromosphere and two orders higher than needed to heat the quiet corona \citep{Withbroe1977}. Already a 10\% wave energy dissipation into heat would be sufficient to heat the chromosphere and corona. Further investigations based on soon available \texttt{ANTARES} RMHD (radiation magnetohydrodynamics) photospheric models will be crucial to test these first assumptions and to determine modifications due to the involvement of the photospheric magnetic field.

Correlation analysis of characteristic parameters, which is the focus of the present preliminary study, is an approved method for examining the solar granulation dynamics and photospheric stratification and dates back to as early as the 1950s, when e.g. \citet{Stuart1954} found a correlation between velocity and brightness---a result that pointed to the no longer surprising fact that bright granules are associated with convective matter upflow. \citet{Durrant1982} found that this correlation extends from the continuum to a height of 300~km beyond that layer, corresponding to a ``pure convective component'' of the velocity field scaling about $3''$ in size, while the velocity field on smaller scales appeared to be more turbulent. Height dependent correlations were analyzed by \citet{Gadun2000} and references listed therein to study the vertical photospheric structure. The former found the top convection zone to reach 20 to 50~km into the photosphere, from where an onset of convective overshoot in stable regions into a height of 150 to 170~km was observed. Beyond that zone the columnar photospheric structure is still maintained due to the influence of convective pressure variations up to a height of about 250 to 300~km, a zone referred to as transition layer and being characterized by its pronounced inversion of temperature fluctuations. A breakdown of the columnar structure takes place at a height of 300~km, from where oscillations govern the dynamics of the photospheric medium. These findings based on coherence analysis confirmed prior results from spectral observations \citep{Nesis1988,Karpinsky1990}.

The remainder of this paper is organized as follows: In Sect.~\ref{sec:model_description} the fundamental equations of radiation hydrodynamics are introduced. We further outline important numerical methods and discuss their advantages for the reliability of the model results. Also boundary and initial conditions as
applied in the particular model presented here are addressed. In Sect.~\ref{sec:data_analysis} some basic techniques for the analysis of the simulation data are presented. For the study of the photospheric structure we first examine the height variation of typical model quantities' horizontal distributions in Sect.~\ref{sec:stratification} and discuss some of their correlations directly at the solar surface. Relative fluctuations of the temperature, gas pressure, and density give important insights into the diverse dynamics of the different photospheric layers and are discussed in Sect.~\ref{sec:rel_fluctuations}. A fuller picture of the photospheric stratification is gained from studying the correlation of characteristic model quantities locally as well as by a two-point correlation, where one quantity is fixed at the surface while the other one is varied with height, which is the objective of Sect.~\ref{sec:local_and_2point_correlations}. Finally, in Sect.~\ref{sec:discussion} we discuss the model results and present possibilities for further investigation.

\section{Methods}
\subsection{Model description}\label{sec:model_description}
The study at hand is based on a simulation of the photospheric matter by the code \texttt{ANTARES} \citep{Muthsam2007,Muthsam2010}. For the quiet Sun the dynamics of the solar surface layers can be described by the equations of radiation hydrodynamics, i.e. the continuity equation
\begin{equation}
\label{eqn:continuity_eqn} \frac{\partial\varrho}{\partial t} + \nabla\cdot(\varrho\boldsymbol{u}) = 0,
\end{equation}
Euler's equation of momentum balance
\begin{equation}
\frac{\partial\varrho\boldsymbol{u}}{\partial t} + \nabla\cdot (\mytensor{M} - \boldsymbol{\sigma}) = \boldsymbol{f},
\end{equation}
and an energy balance
\begin{equation}
\label{eqn:energy_conservation} \frac{\partial e}{\partial t} + \nabla\cdot\bigl(\boldsymbol{u} (e + P) - \boldsymbol{u} \cdot\boldsymbol{\tau} \bigr) = \varrho(\boldsymbol{g} \cdot \boldsymbol{u}) + Q_\mathrm{rad},
\end{equation}
complemented by the equation of state closing the conservation laws (\ref{eqn:continuity_eqn}) to (\ref{eqn:energy_conservation}).\footnote{\texttt{ANTARES} takes
into account partial ionization and realistic microphysics by adopting the OPAL equation of state \citep{Iglesias1996}. It is based on Rosseland opacities by \citet{Iglesias1996} and by \citet{Alexander1994,Ferguson2005} at low temperatures in the case of gray radiative transfer. For the non-gray radiative transfer opacity distribution functions and model atmospheres of the \texttt{ATLAS 9} package of \citet{Kurucz1970} are used. In Fig.~\ref{fig:vertical_runs} the adopted model atmosphere of \citet{Kurucz1970} is compared to the energy-balance model atmosphere of the quiet Sun of \citet{Fontenla1993}.} Here $\varrho$ denotes the mass density, $\boldsymbol{u} = u \boldsymbol{e}_x + v \boldsymbol{e}_y + w \boldsymbol{e}_z$ is the flow velocity, $\boldsymbol{\sigma} = - P \mytensor{I} + \boldsymbol{\tau}$ is the stress tensor
with the kinetic gas pressure $P = -\nicefrac{1}{3}\operatorname{tr}\boldsymbol{\sigma}$
and the viscous stress tensor $\boldsymbol{\tau} = \mu ( \nabla\boldsymbol{u} + (\nabla
\boldsymbol{u})^\mathrm{T} ) + \nicefrac{2}{3} \mu\nabla\cdot \boldsymbol{u} \mytensor{I}$, where $\mu$ is the dynamic molecular viscosity. Furthermore $\boldsymbol{f} = \varrho\boldsymbol{g}$ denotes an external force density with $\boldsymbol{g}$, the local gravitational acceleration, $\mytensor{M} = \varrho \boldsymbol{u} \otimes\boldsymbol{u}$ is the momentum current density tensor, $e = \nicefrac{1}{2}
\varrho\boldsymbol{u}^2 + \varrho \mathscr{E}$ is the total kinetic energy density made up of the convective kinetic- and the internal energy density, and $Q_\mathrm{rad}$ represents radiative sources.

In the scope of RHD finally the frequency-dependent and time-independent\footnote{As travel times of photons through the photosphere are much shorter than any other time scales involved.} radiative transfer has to be considered. The radiation transfer equation
\begin{equation}
\hat{\boldsymbol{r}} \cdot\nabla I_\nu= \varrho\kappa_\nu (S_{\nu} - I_\nu)
\end{equation}
with spectral intensity $I_\nu$, source function $S_{\nu}$, and material opacity $\kappa_\nu$ is solved for the upper $\sim1~\mbox{Mm}$ of the computational domain. It is linked to the energy balance equation via the radiative heating rate $Q_\mathrm{rad}= - \int(\nabla\cdot \boldsymbol{F}_\nu) \, \mathrm{d} \nu$, accounting for the energy exchange between the gas and the radiation field, where $\boldsymbol{F}_\nu$, the frequency $\nu$-dependent radiative energy flux is the spectral intensity $I_\nu $ integrated over the solid angle $\varOmega$ into that energy is radiated along unit vector $\hat{\boldsymbol{r}}$, i.e. $\boldsymbol{F}_\nu= \int I_\nu(\hat{\boldsymbol{r}}) \hat{\boldsymbol{r}}\,\mathrm{d} \varOmega$. By solving the equations of radiation-hydrodynamics, the code simulates convection and full (i.e. non-gray) radiative transfer in local thermal equilibrium (LTE).\footnote{Whereby the source function is considered equal to the Planck function $S_{ \nu} = B_\nu(T)$.} The full radiative flux $\boldsymbol {F}_\mathrm{rad}$ is found by integrating $\boldsymbol{F}_\nu$ over frequency using non-gray opacities with $N=4$ bins, applying the frequency binning method developed
originally by \citet{Nordlund1982} and used extensively thereafter. At a sufficient depth, typically $\tau \gtrsim100$ \citep{Steiner1997}, the heating rate can be computed according to the diffusion approximation $Q_\mathrm{rad} = \nabla\cdot(\kappa\nabla T)$.

The temporal integration of the model equations uses second- or third-order Runge-Kutta methods with weighted essentially non-oscillatory (WENO) finite volume schemes. The code is based on a high-resolution finite-volume method that can treat turbulence by adopting local mesh refinement. Essentially, finite volume schemes are based on interpolation of discrete data using polynomials; fixed stencil interpolations work well for sufficiently smooth problems but introduce oscillations near discontinuities, whose amplitudes do not
decay in the course of mesh refinement. Whereas traditional remedies such as the introduction of an artificial viscosity or the application of limiters to discard such oscillations have obvious drawbacks, ENO (originally developed by \citealp{Harten1987,Harten1987b}) and Weighted ENO schemes\footnote{Which, in contrast to the ENO scheme uses a convex combination of all candidate stencils.} \citep{Liu1994,Jiang1996} are based on a nonlinear adaptive procedure to determine the locally smoothest stencil, whereby the crossing of discontinuities in the interpolation procedure can be avoided by and large \citep{Shu1998}. Further information concerning the implementation of WENO in \texttt{ANTARES} can be found in \citet{Kupka2012,Mundprecht2013,Zaussinger2013}, and \citet{GrimmStrele2014}.

In horizontal directions, periodic boundary conditions for all quantities are applied. The current model has however been improved lately w.r.t. the applied vertical boundary conditions. In principle, boundary conditions should influence the flow kinematics at the boundaries as little as possible. By originally inhibiting any vertical convective matter and energy flux through the lower and upper boundaries by setting the vertical flow velocity component to zero, $u|_\mathrm{top} = u|_\mathrm{bot} = 0$, mass and energy conservation had been ensured. While such closed boundary conditions have the advantage of simplicity and high stability, they disturb the velocity field in an undesirable way \citep{Robinson2003} and are in general prone to non-physical reflections of waves and shocks and had been justified as the optical depth unity layer had only been weakly influenced \citep{Mundprecht2013}. Also the energy fluxes have been found to be influenced by forcing zero fluid velocity to an extent of about two pressure scale heights above the lower end of the computational domain \citep{GrimmStrele2015}. The current model remedies these deficiencies by applying open boundary conditions at the top and at the bottom, thereby allowing for convective mass and energy in- and outflows. This is of special significance at the lower boundary where most of the energy transport is due to advection and kinetic energy flux. The lack of convective matter
flux into the bottom layer in prior models had to be compensated for by an appropriate artificial radiative source term. The boundaries are made transmissive for waves by constantly extrapolating velocities according to
\begin{equation}
\frac{\partial u}{\partial x} = \frac{\partial v}{\partial x} = \frac {\partial w}{\partial x} = 0,
\end{equation}
thereby increasing the long-term stability of the model \citep{GrimmStrele2015}. The latter study gives a detailed description of the recent numerical implementations of
boundary conditions in the \texttt{ANTARES} code and also compares the effects of closed and several open boundary conditions. 

For initial conditions a solar atmosphere between layers of $4350~\mbox{K}$ at the top and $20\,000~\mbox{K}$ at the bottom, respectively is considered. The horizontal momentum densities $\varrho v$ and $\varrho w$ are slightly disturbed before the system is evolved in time.

The simulation time exceeds 5~hours with a time step of $\Delta t = 8.7~\mbox{s}$ between snapshots. The model domain is covered by a Cartesian lattice $\varOmega= \{(i,j,k)|i,j,k \in\mathbb{N}_0; 0 \leq i \leq404; 0 \leq j,k \leq510\}$ ranging from the top of the photosphere to $\sim4~\mbox{Mm}$ into the upper convection zone and covering a field of $18~\mbox{Mm} \times18~\mbox{Mm}$ in horizontal direction. The considerably high spatial horizontal resolution of $\Delta y = \Delta z = 35.3~\mbox{km}$ exceeds the resolution of recent observational data, e.g. images of the 1~m~Swedish Solar Telescope (SST) that, neglecting the loss of resolution capacity due to seeing conditions, has a theoretical diffraction limit of e.g. $\lambda/D \approx
0.14 ''$ in $\mbox{H}\alpha$ \citep{Antolin2012}, corresponding to $\sim 100~\mbox{km}$ on the solar disc, but also of the 1.6~m~Solar Telescope at Big Bear Observatory with a diffraction limit of $\approx0.07''$ at 500~nm \citep{Goode2002} and of the 1.5~m~GREGOR telescope at the Observatorio del Teide, Tenerife which resolves spatial scales on the solar surface down to $\approx 50~\mbox{km}$ \citep{Puschmann2012}. The vertical dimension of the model grid is even better resolved with $\Delta x \approx 11.0~\mbox{km}$, enabling us to accuratelyexamine the photospheric stratification. The top of the grid is associated with zero optical depth. With increasing opacity, the solar surface is found to be located on average $\approx 600~\mbox{km}$ below.

\subsection{Data analysis techniques}\label{sec:data_analysis}
A first insight into the structure of the solar photosphere is gained from analyzing relative fluctuations $\delta Q = (Q - \langle Q \rangle)/\langle Q \rangle$ of a quantity $Q$, such as the thermodynamic state variables temperature $T$, gas pressure $P$ and mass density $\varrho$. Here $\langle Q \rangle$ denotes a horizontal average of the particular quantity in question, evaluated separately in upflow- $U = \{(i,j,k)|u_{i jk} \leq0\}$ and downflow regions $D = \{(i,j,k)|u_{ijk} > 0\}$ with vertical flow velocity $u = \boldsymbol{u} \cdot\boldsymbol{e}_x$. The vertical profiles of these fluctuations are studied after having been finally averaged horizontally over up- and downflow regions and over time.

The linear dependency between two matrix-valued quantities $A_{jk}$ and $B_{jk}$ evaluated at horizontal surfaces for a given vertical level $i$ is described by the linear correlation coefficient
\begin{equation}
\label{eqn:corr_coeff} \rho_i = \frac{\sum_{j,k} (A_{i\!jk} - \bar A_i ) (B_{l\!jk} - \bar B_l )}{\sqrt{ (\sum_{j,k} ( A_{i\!jk} - \bar A_i )^2 ) ( \sum_{j,k} (B_{l\!jk} - \bar B_l )^2 )}},
\end{equation}
which for $l = i$ measures the linear dependence at the same depth, also referred to as local one-point correlation. In general, $l \neq i$, which is also referred to as two-point correlation and from which changes in the columnar photospheric structure can be analyzed \citep{Gadun2000}. Note that e.g. $\bar A_i$ simply denotes the horizontal mean of $A_{i\!jk}$ for fixed $i$. Both kinds of methods are used extensively in the following to measure the respective local and non-local correlations between fluctuations of temperature $\delta T$, pressure $\delta P$, mass density $\delta\varrho$, opacity $\delta\kappa$ and vertical fluid flow component $\delta u$.

\begin{figure}[htb!]
\includegraphics[width=0.45\textwidth]{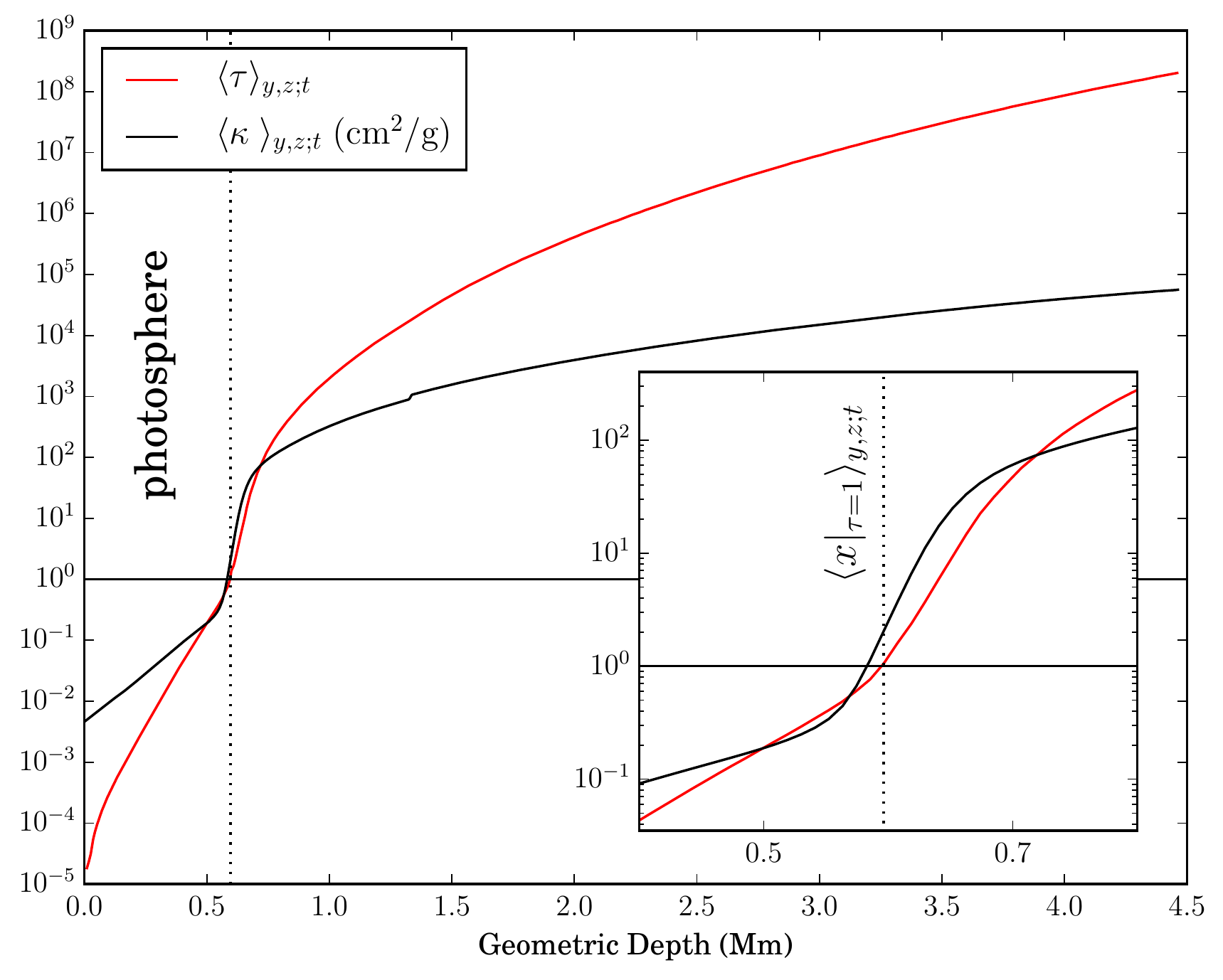}
\caption{Profile of the temporally and spatially averaged optical depth $\tau(x)$ as a function of geometric depth as obtained by numerical integration of Eq.~(\protect\ref
{eqn:tau1})} \label{fig:tau_kappa_run}
\end{figure}

\begin{figure}[htb!]
\includegraphics[width=\columnwidth]{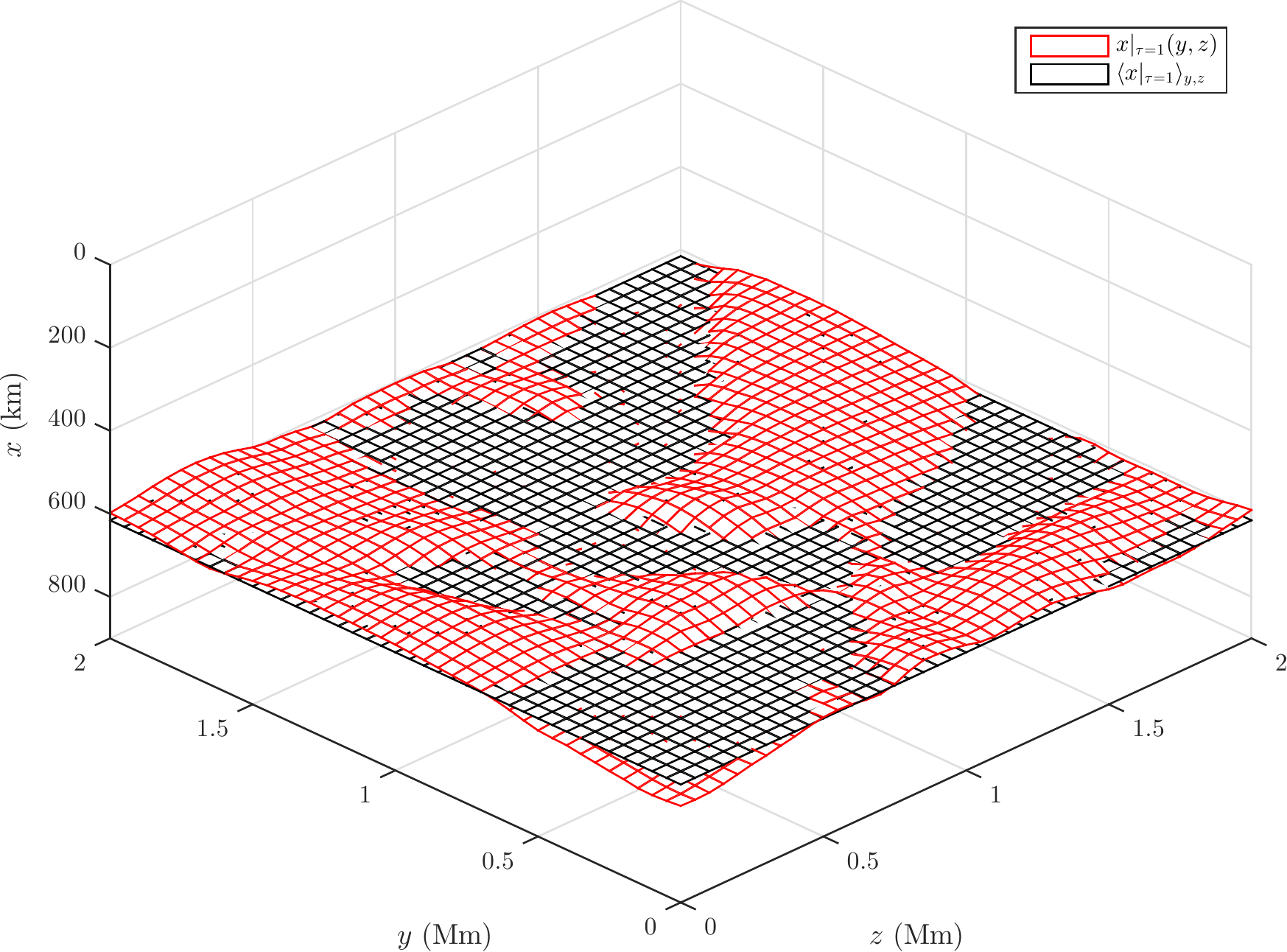}
\caption{Temporal snapshot of $x|_{\tau=1}(y,z)$ (\emph{red}) and the horizontal average $\langle x|_{\tau=1} \rangle_{y,z}$ (\emph{black}). Higher layers, i.e. smaller values of $x$ correspond to the granular brightness field, while impressed regions are associated with the intergranular lanes, where the solar surface is observed at deeper layers} \label{fig:tau1_isosurface}
\end{figure}

\begin{figure*}[htb!]
\includegraphics[width=0.458\textwidth]{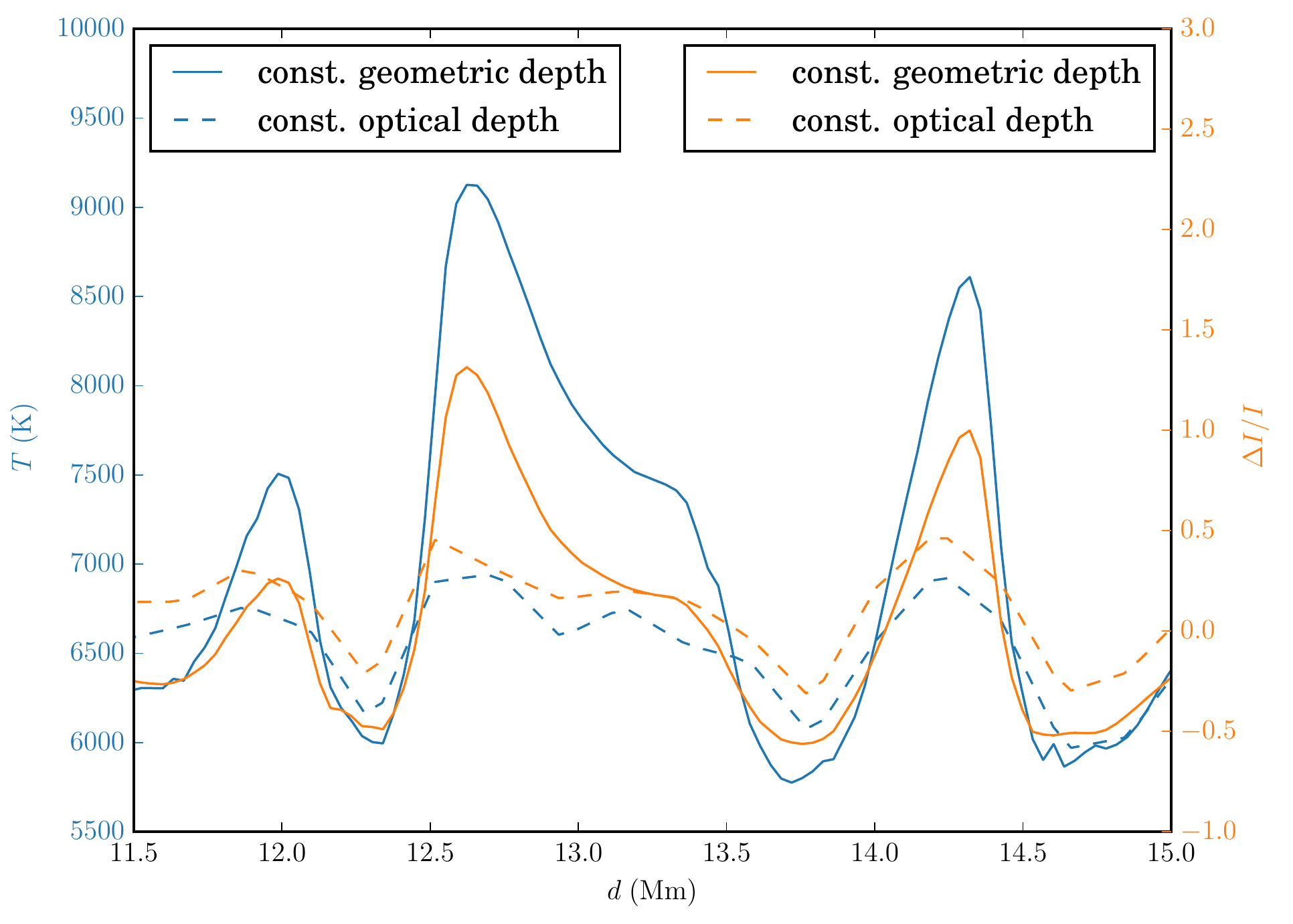}
\includegraphics[width=0.542\textwidth]{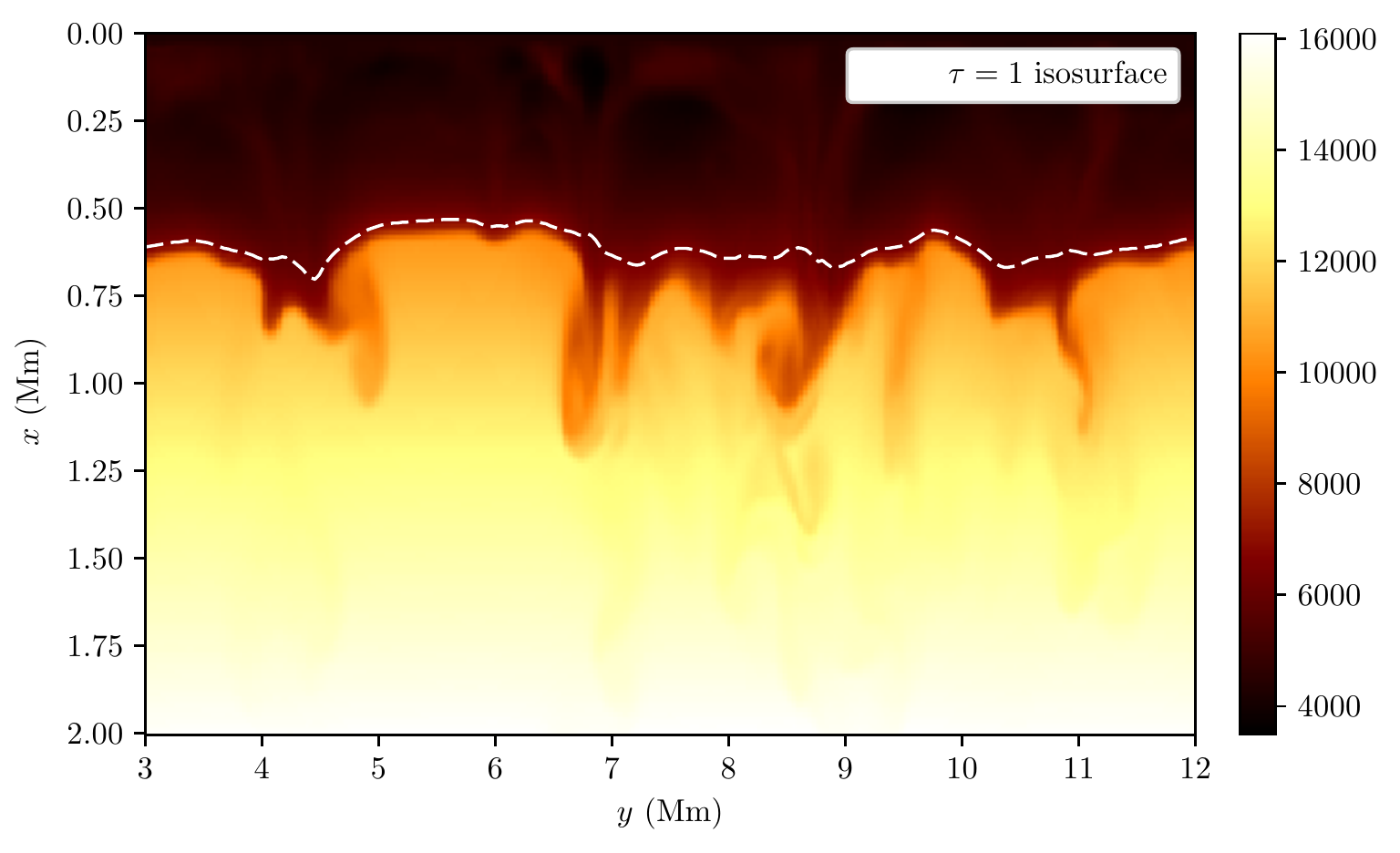}
\caption{\textit{Left:} Temperature and relative intensity runs at a horizontal slice at constant geometric depth corresponding to the horizontal mean of optical depth unity (solid) and along a slice lying at the $\tau=1$ isosurface (\emph{dashed}). \textit{Right:} Temperature distribution $T(x,y)$ (K) and $\tau=1$ isosurface at a vertical cross section\hfill\break} \label{fig:T_distribution}
\end{figure*}

The solar surface corresponding to $\ln\tau= 0$, i.e. optical depth unity, serves as a reference level for the two-point correlations defined above. The boundary at the upper vertical level of the grid $x_i = 0$ is associated with a zero optical depth $\tau_{0,jk} = 0\ \forall j,k$, from where by a post-processing numerical integration top-down a columnar estimate for the optical depth is obtained,
\begin{equation}
\label{eqn:tau1} \tau_{i jk} \approx\tau_{i-1,jk} + \int_{x_{i-1}}^{x_i} \kappa_{jk}(x) \varrho_{jk}(x) \, \mathrm{d} x \quad (i \geq1).
\end{equation}
Figure~\ref{fig:tau_kappa_run} shows the logarithmic run of $\langle \tau\rangle$ with geometric depth, indicating a maximum optical depth of $\sim10^8$ at the lower end of the lattice inside the convection zone corresponding to a depth of $\approx 3.9~\mbox{Mm}$ below the visible surface. From this estimate, Eq.~(\ref{eqn:tau1}), the mean level of $\tau= 1$ as obtained by averaging over columns and time is found to be located at $\langle x|_{\tau= 1} \rangle \approx 602.8~\mbox{km}$. The evaluated $\tau$-unity isosurface and its horizontal average for a snapshot in a $2 \times2~\mbox{Mm}^2$ subfield are shown in Fig.~\ref{fig:tau1_isosurface}. As the $\mbox{H}^-$ opacity increases with temperature, the corrugated isosurface juts out above the averaged geometric depth level in the hot upflow regions, while in the intergranular lanes deeper layers are observed. The rms-variation of optical depth unity is $\approx 34.8~\mbox{km}$, a value comparable with the optical depth corrugation found from the solar granulation models of \cite{Stein1998}. Since at the constant geometric depth level temperatures in the hot upflow regions are measured deeper down in even hotter layers and higher up in the cooler downflow regions as compared to the visible surface, a much broader temperature- and intensity range is to be expected. This is illustrated in the left panel of Fig.~\ref{fig:T_distribution}, where the temperature profile along a horizontal and a $\tau$-unity slice placed across two minor and a major granule of $\sim1~\mbox{Mm}$ size clearly shows a significantly larger temperature (and accordingly also inten\-sity-) variation when evaluated at the averaged cons\-tant geo\-metric depth level rather than at the $\tau$-unity isosurface itself.

\begin{figure*}
\includegraphics[width=0.245\textwidth]{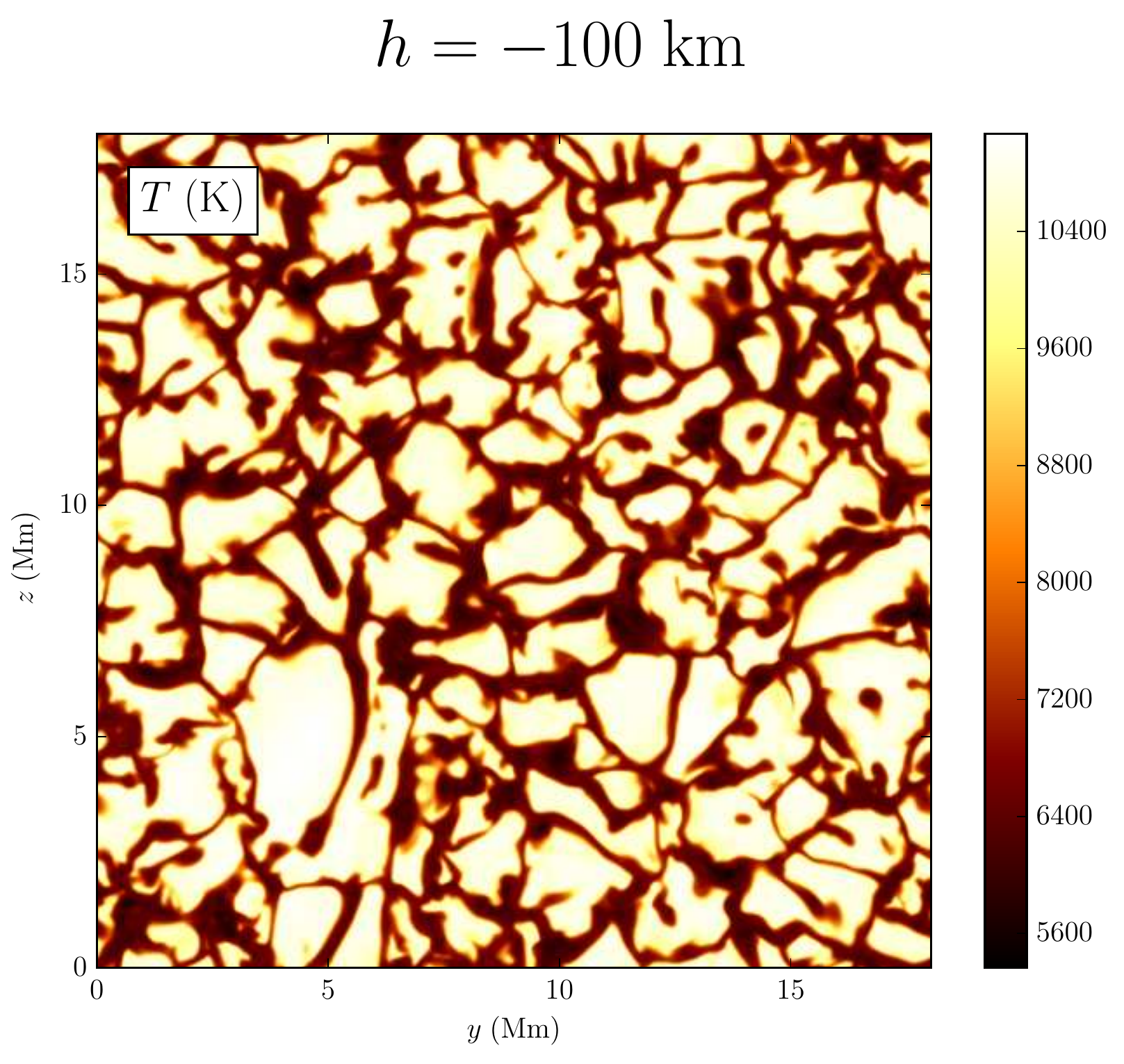}
\includegraphics[width=0.245\textwidth]{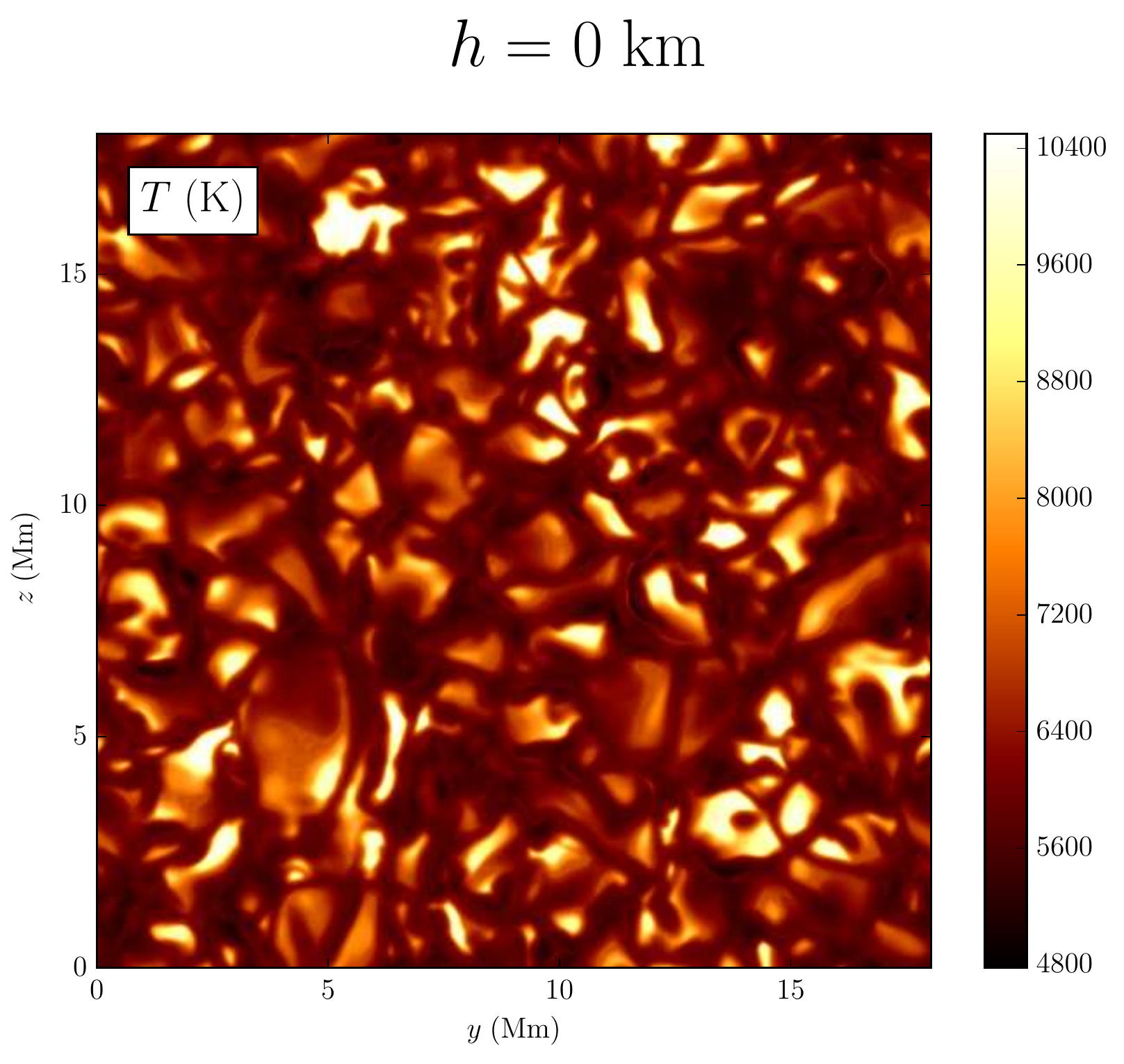}
\includegraphics[width=0.245\textwidth]{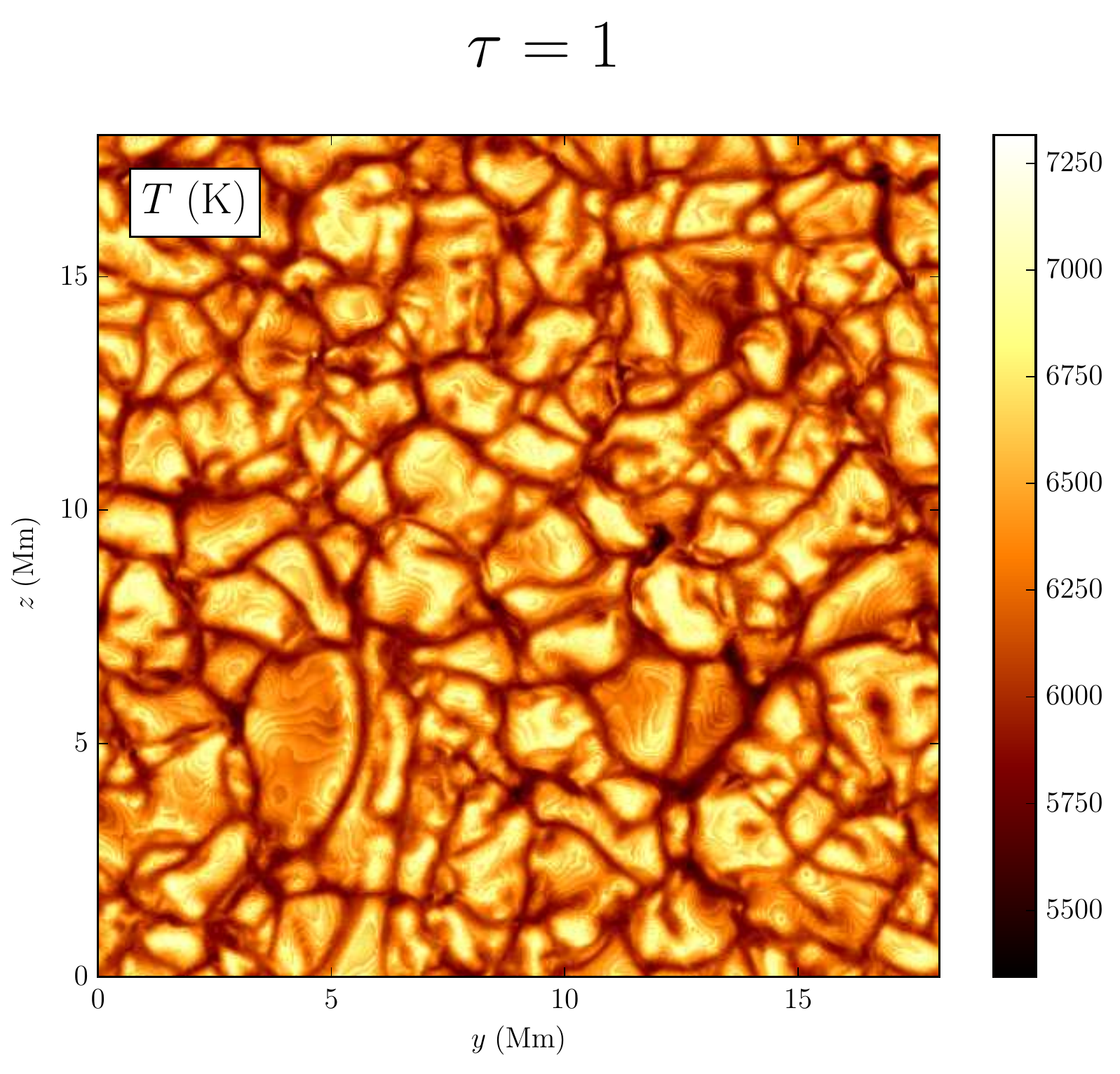}
\includegraphics[width=0.245\textwidth]{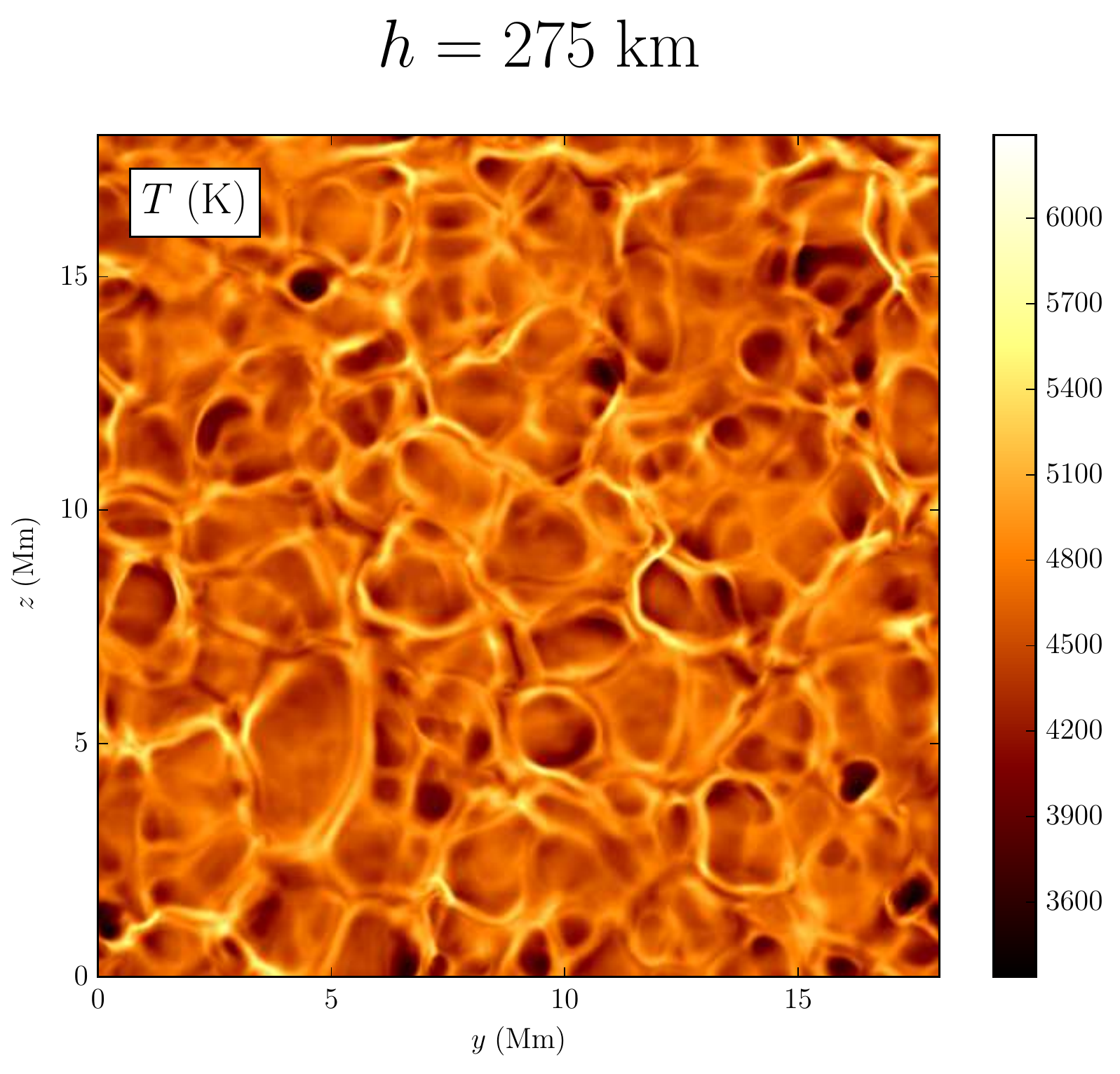}
\includegraphics[width=0.245\textwidth]{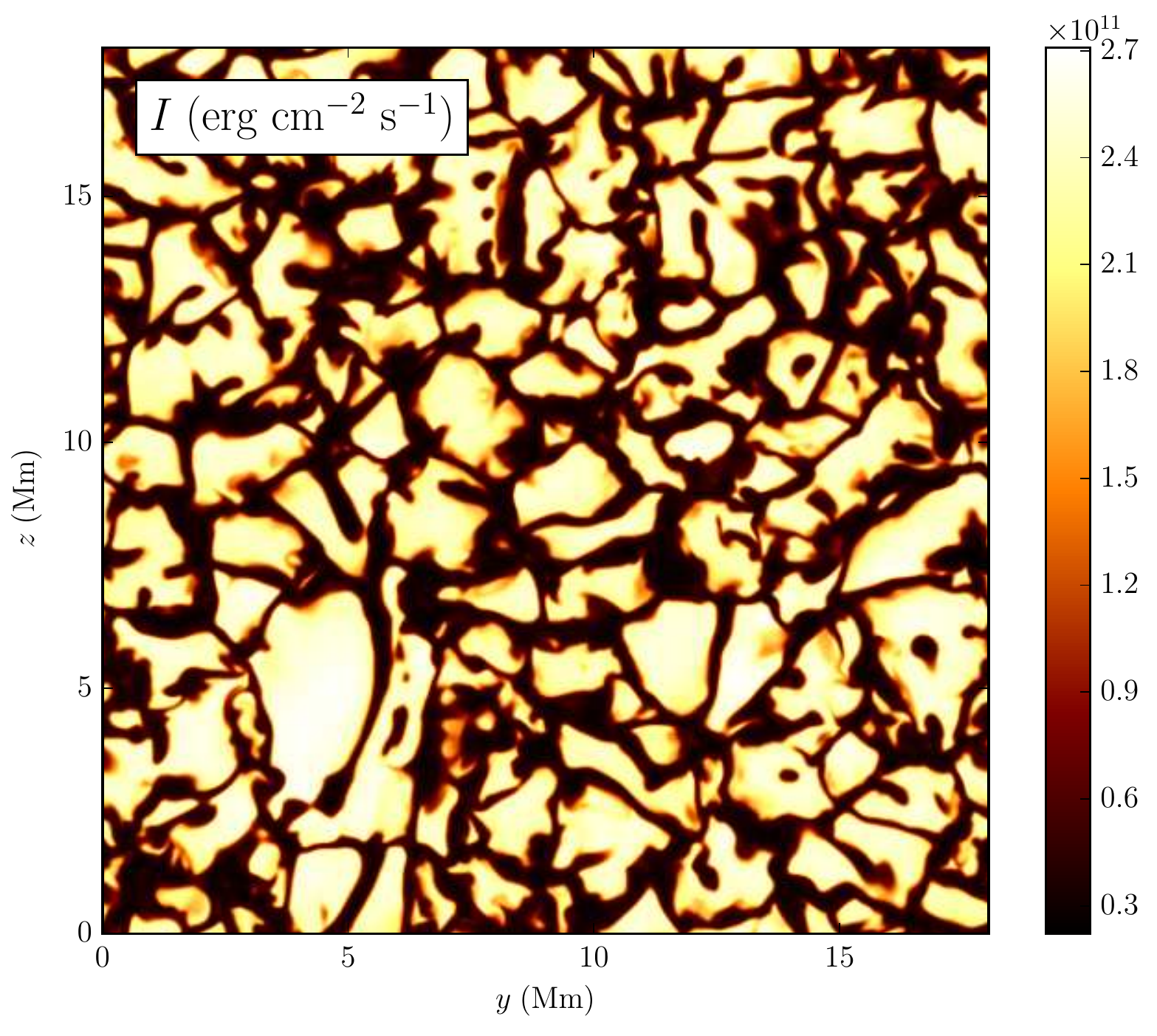}
\includegraphics[width=0.245\textwidth]{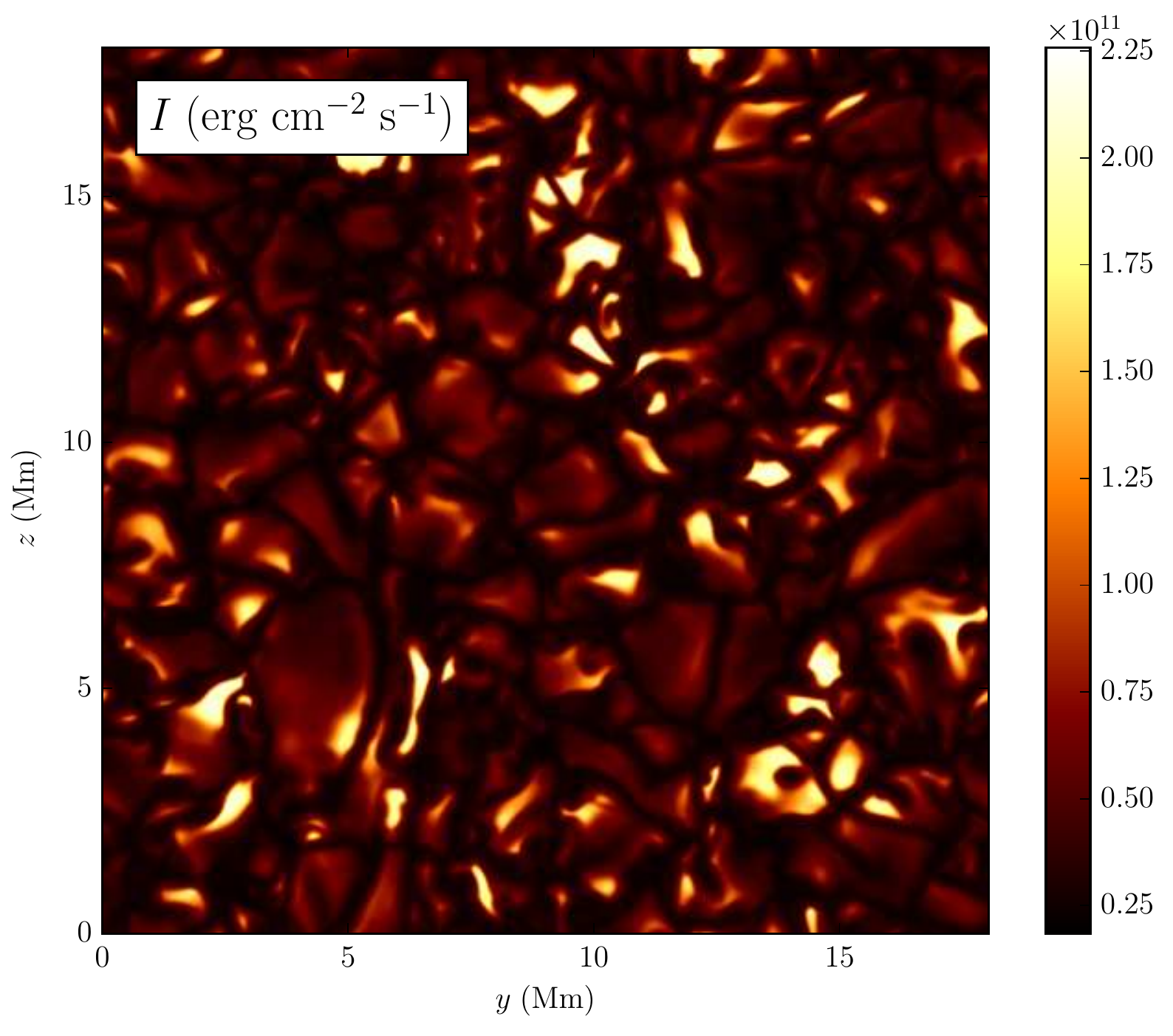}
\includegraphics[width=0.245\textwidth]{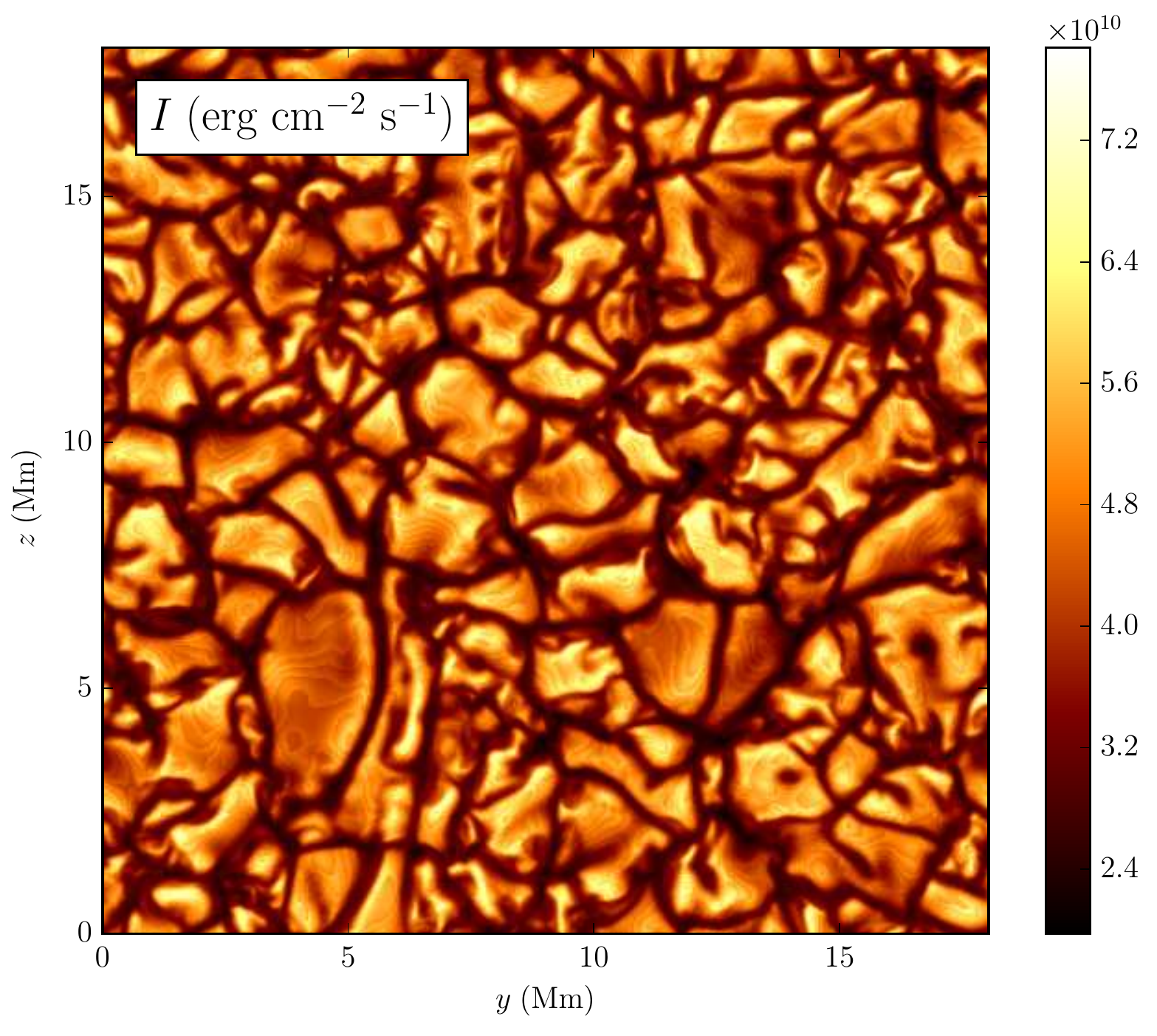}
\includegraphics[width=0.245\textwidth]{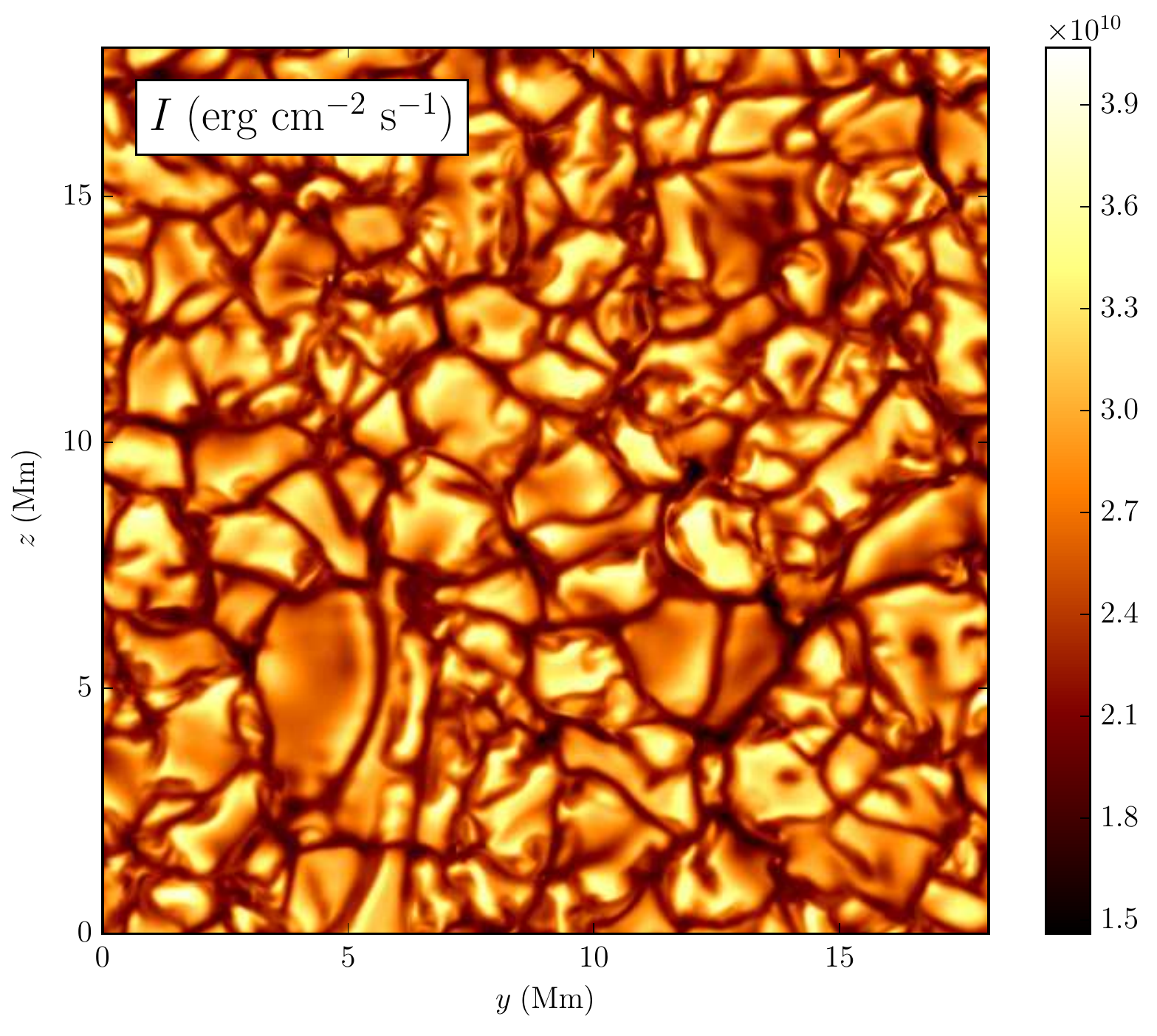}
\includegraphics[width=0.245\textwidth]{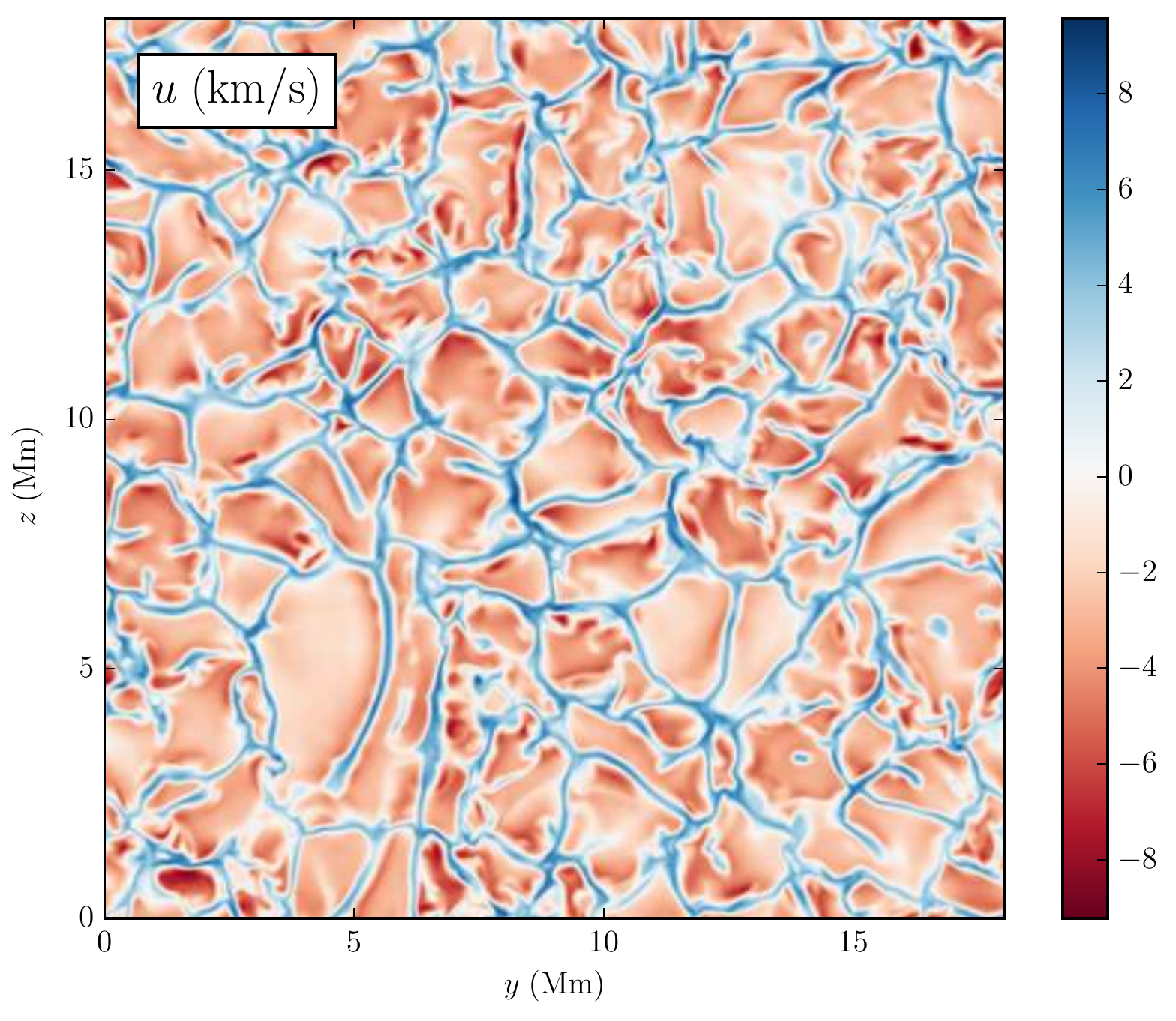}
\includegraphics[width=0.245\textwidth]{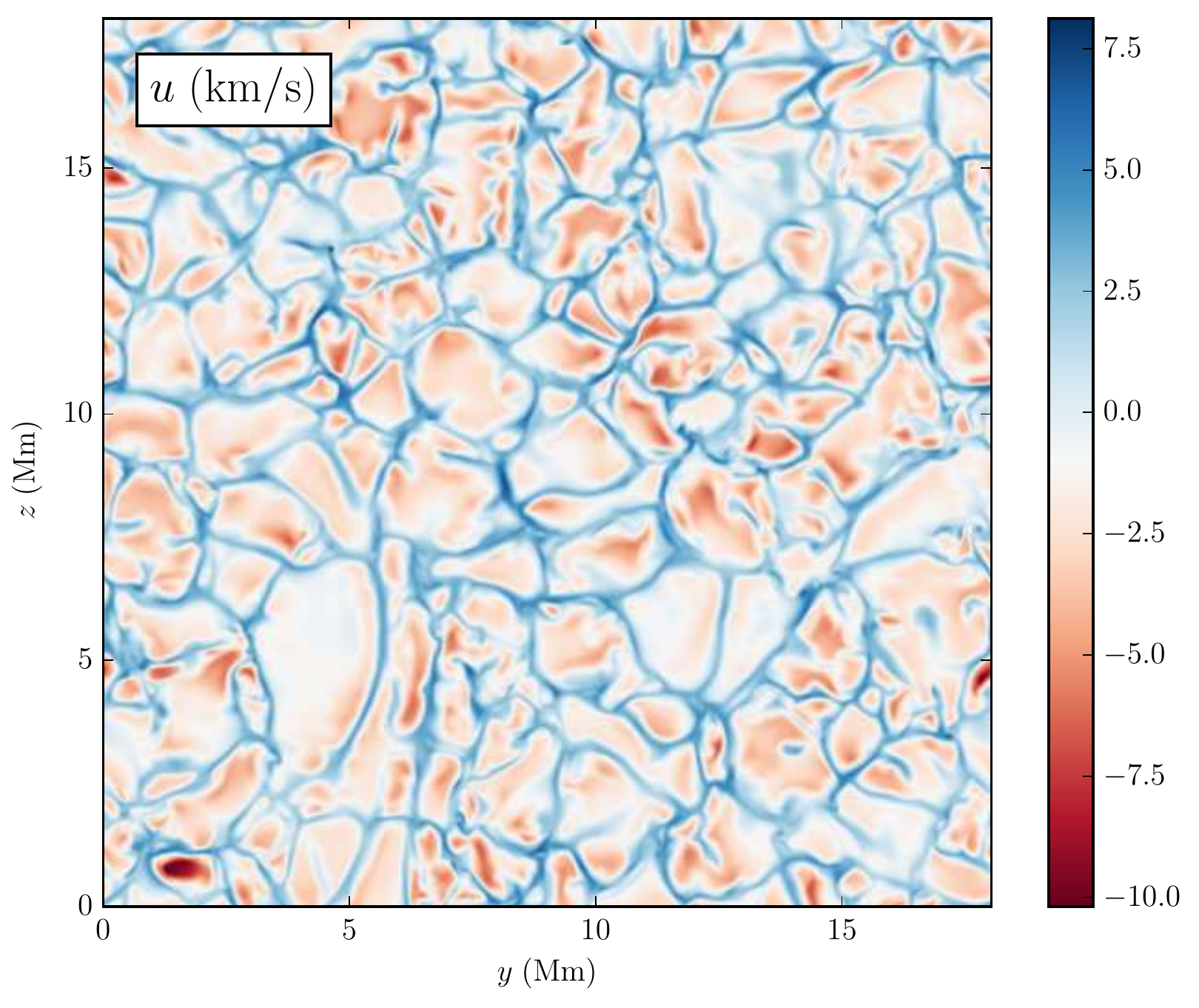}
\includegraphics[width=0.245\textwidth]{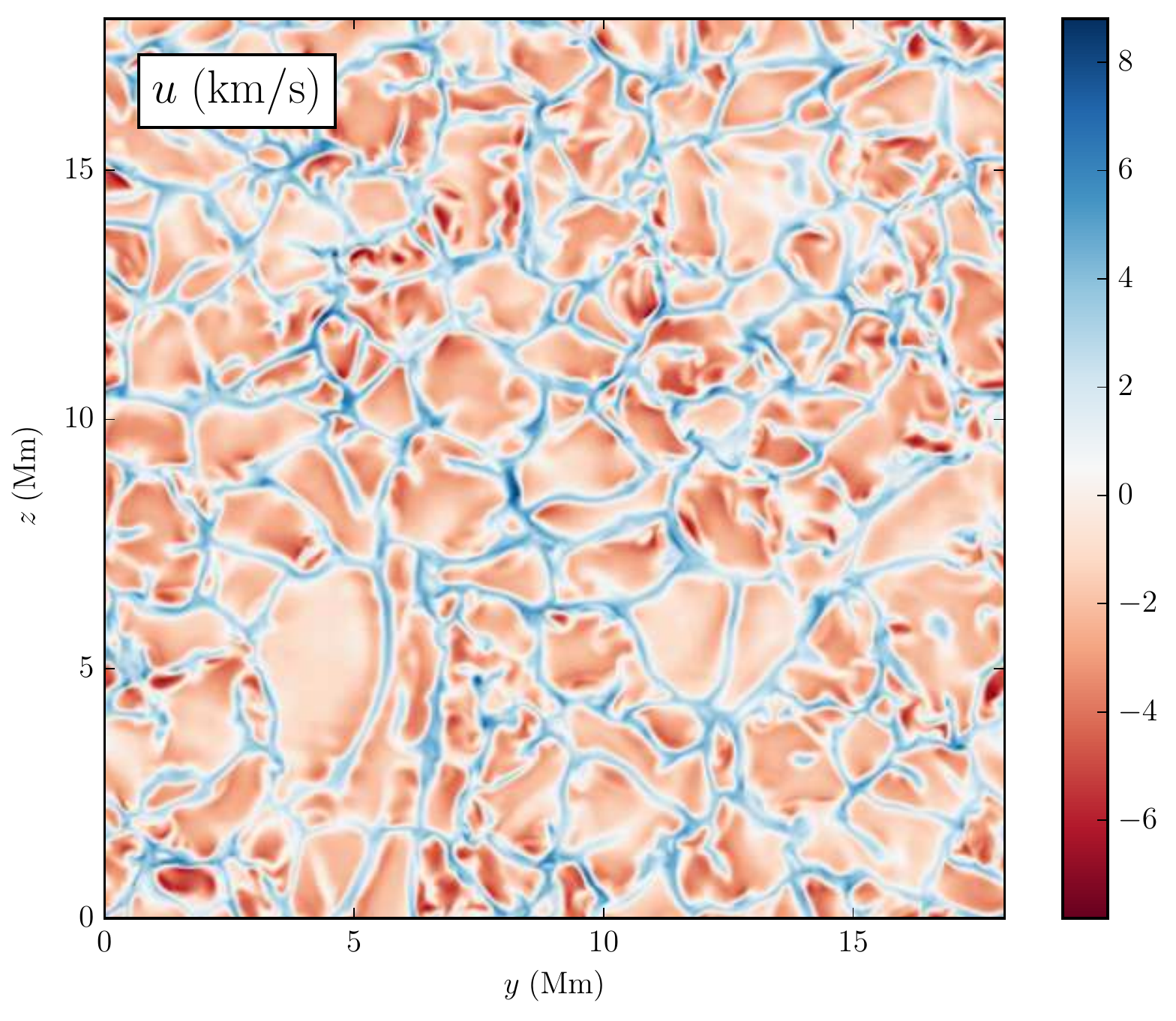}
\includegraphics[width=0.245\textwidth]{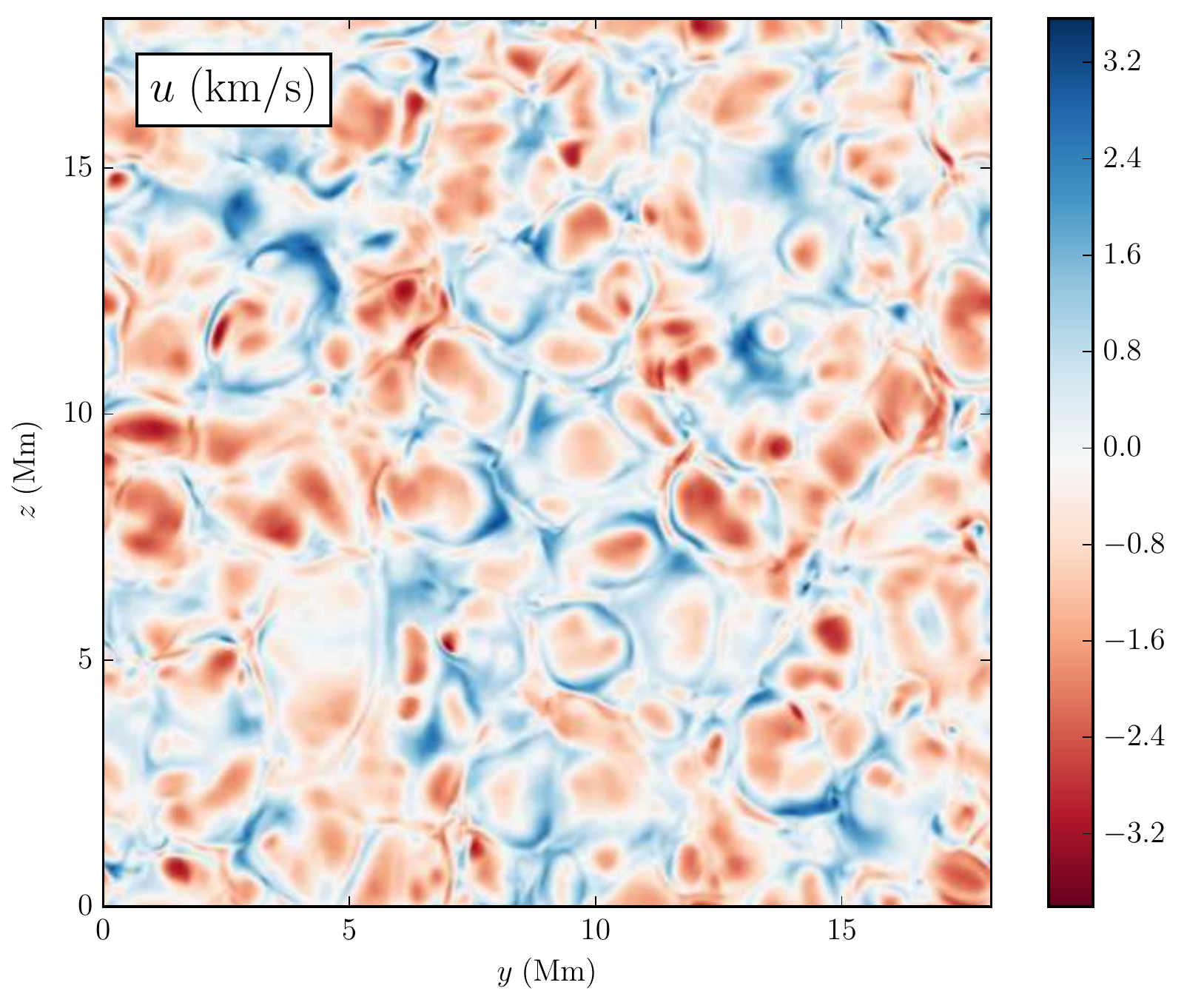}
\includegraphics[width=0.245\textwidth]{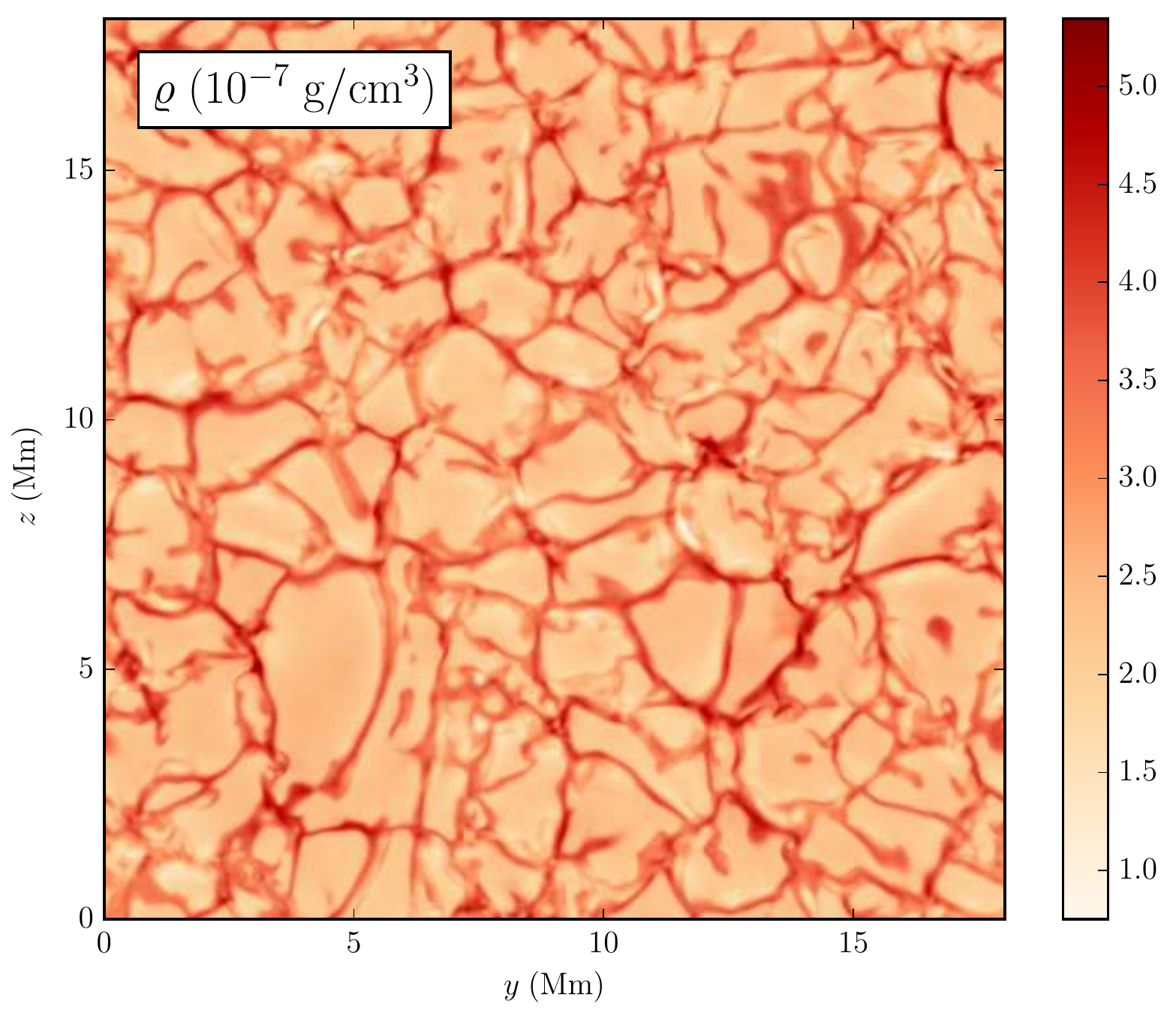}
\includegraphics[width=0.245\textwidth]{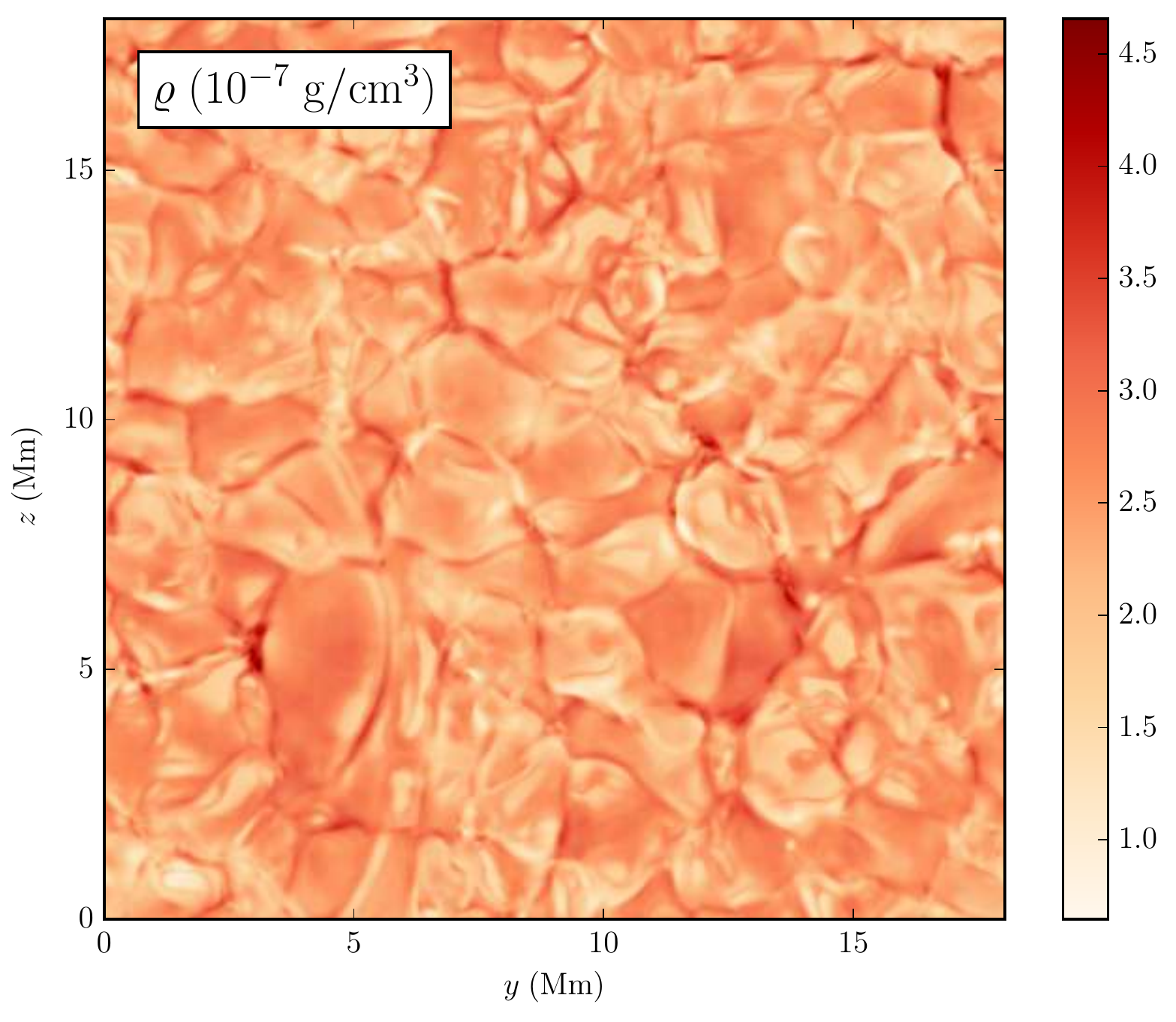}
\includegraphics[width=0.245\textwidth]{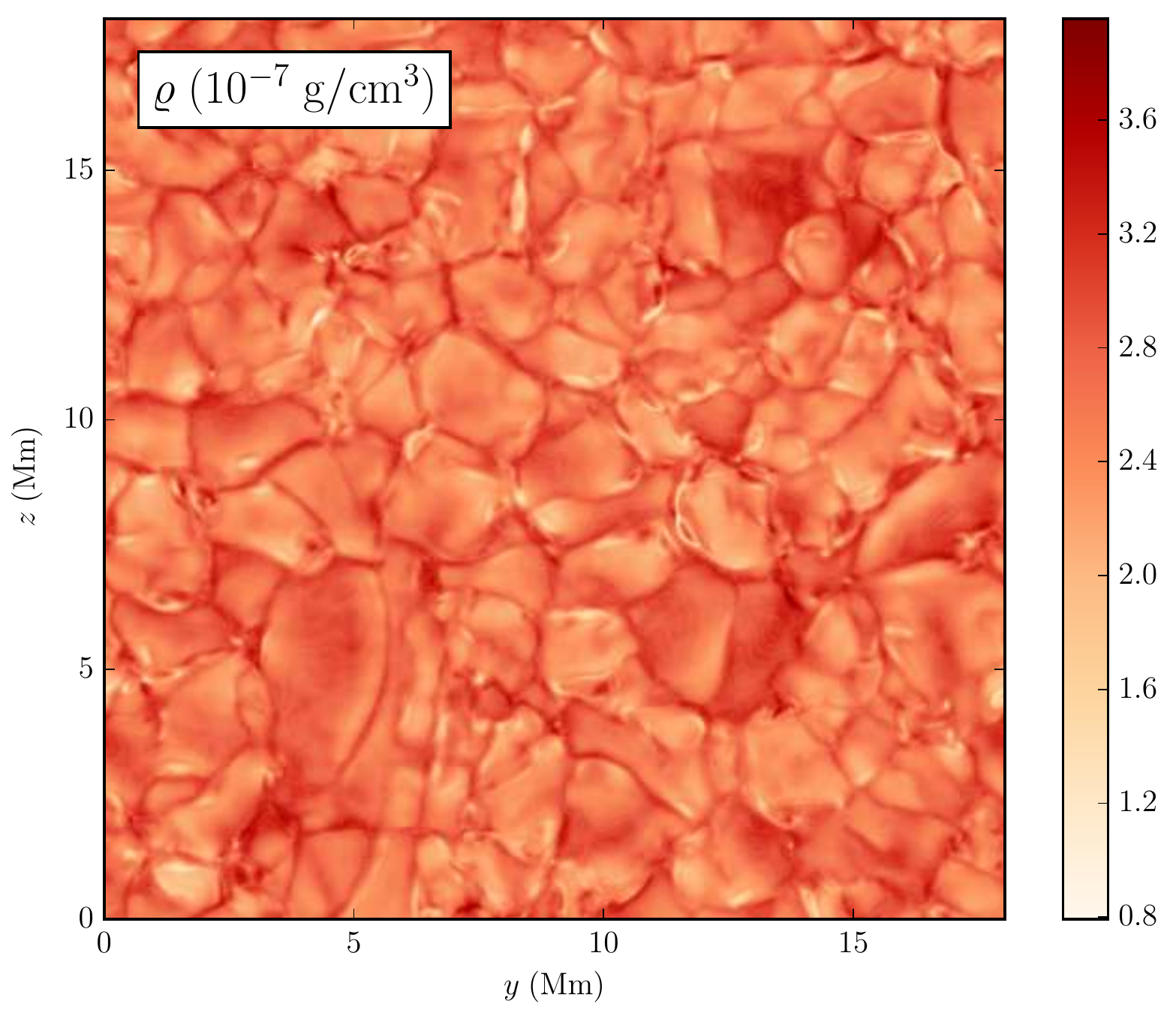}
\includegraphics[width=0.245\textwidth]{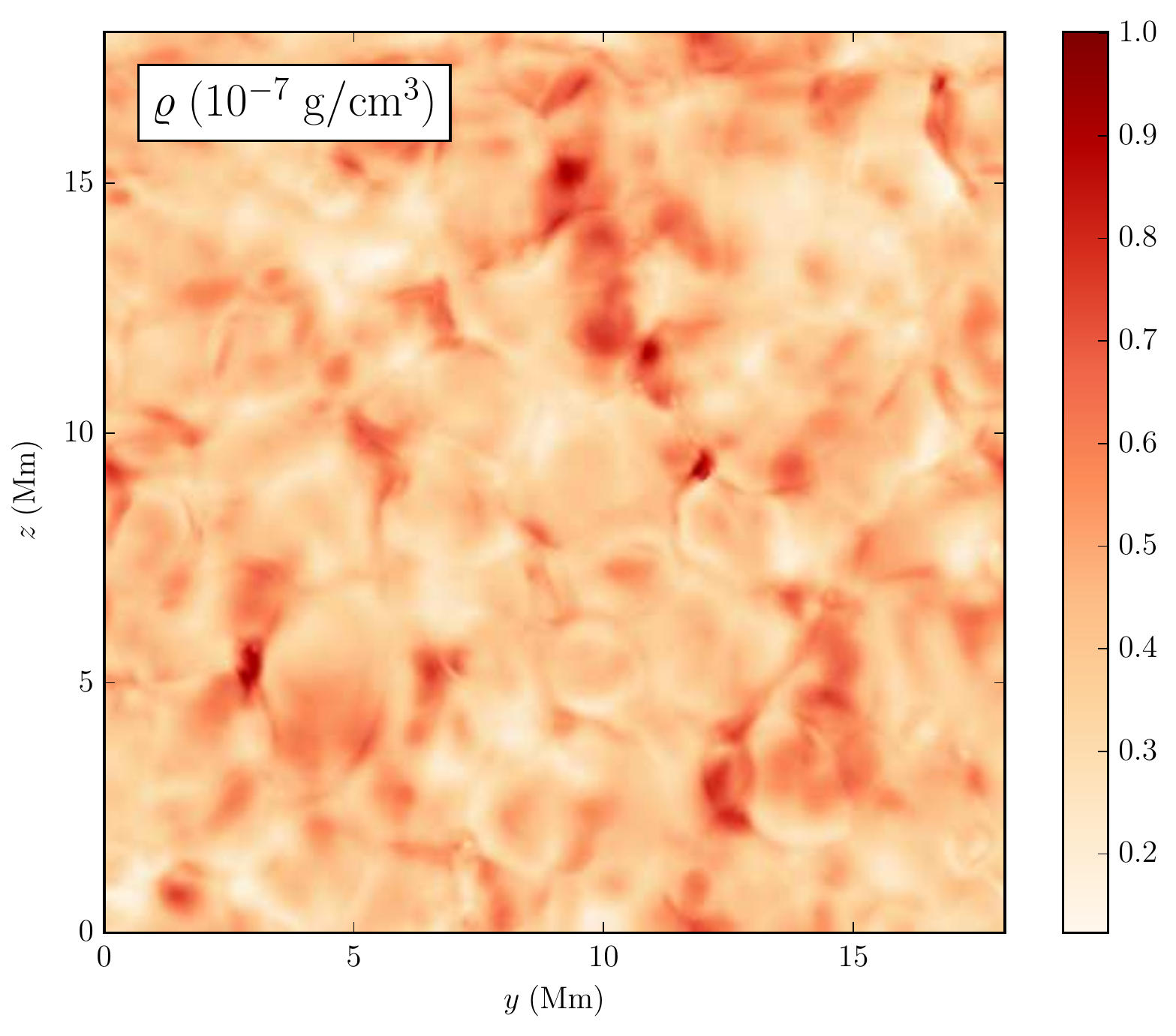}
\includegraphics[width=0.245\textwidth]{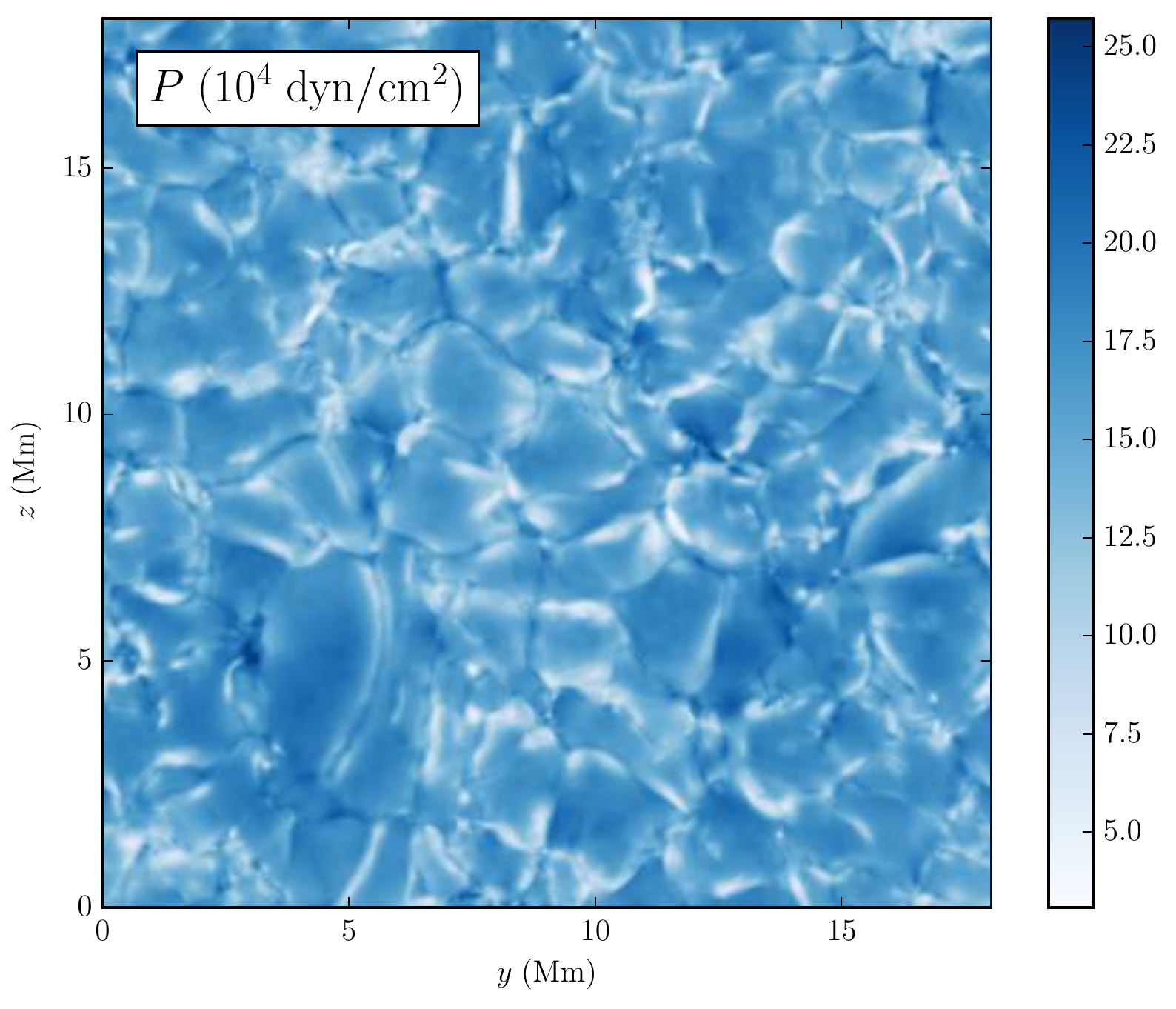}
\includegraphics[width=0.245\textwidth]{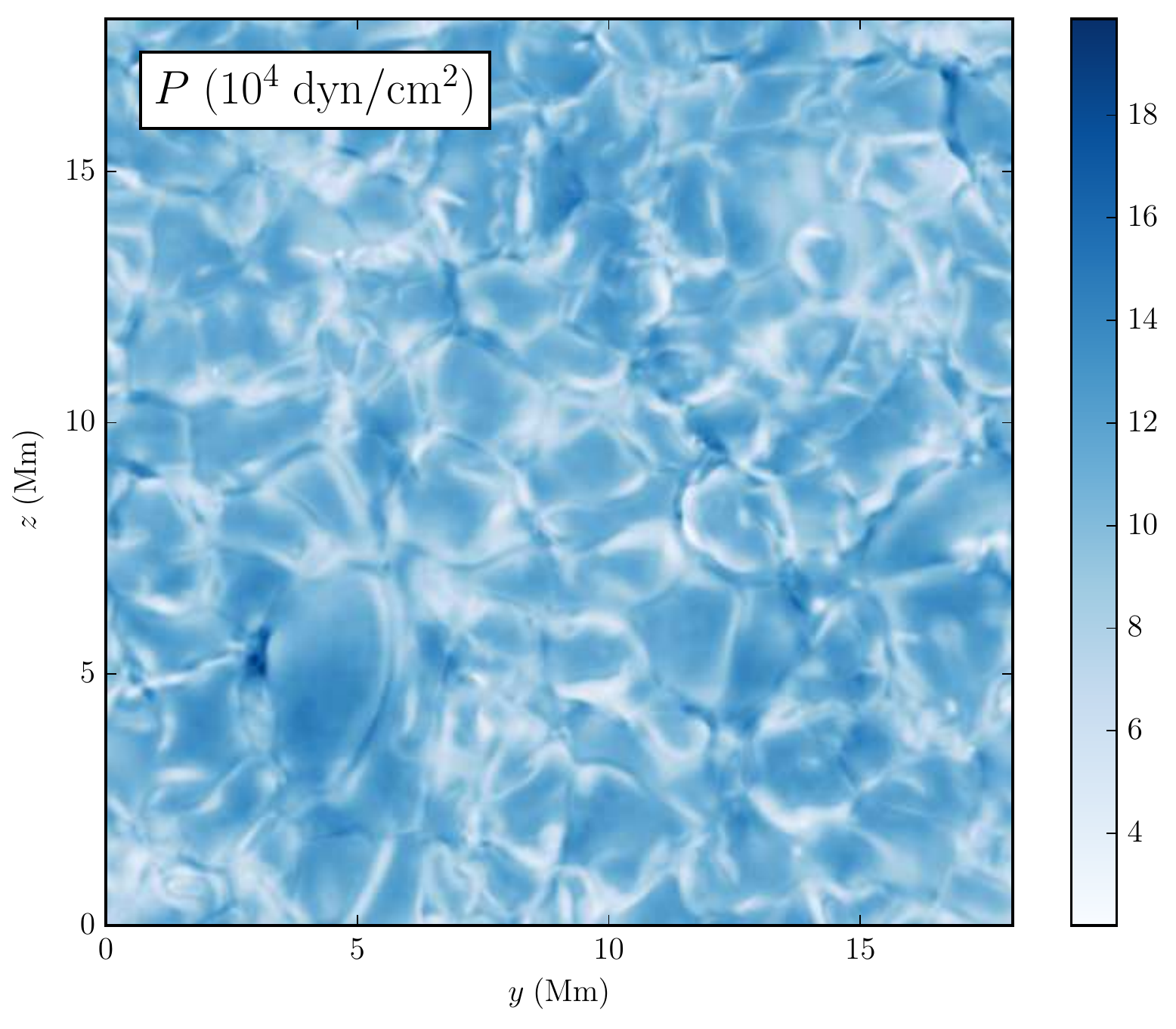}
\includegraphics[width=0.245\textwidth]{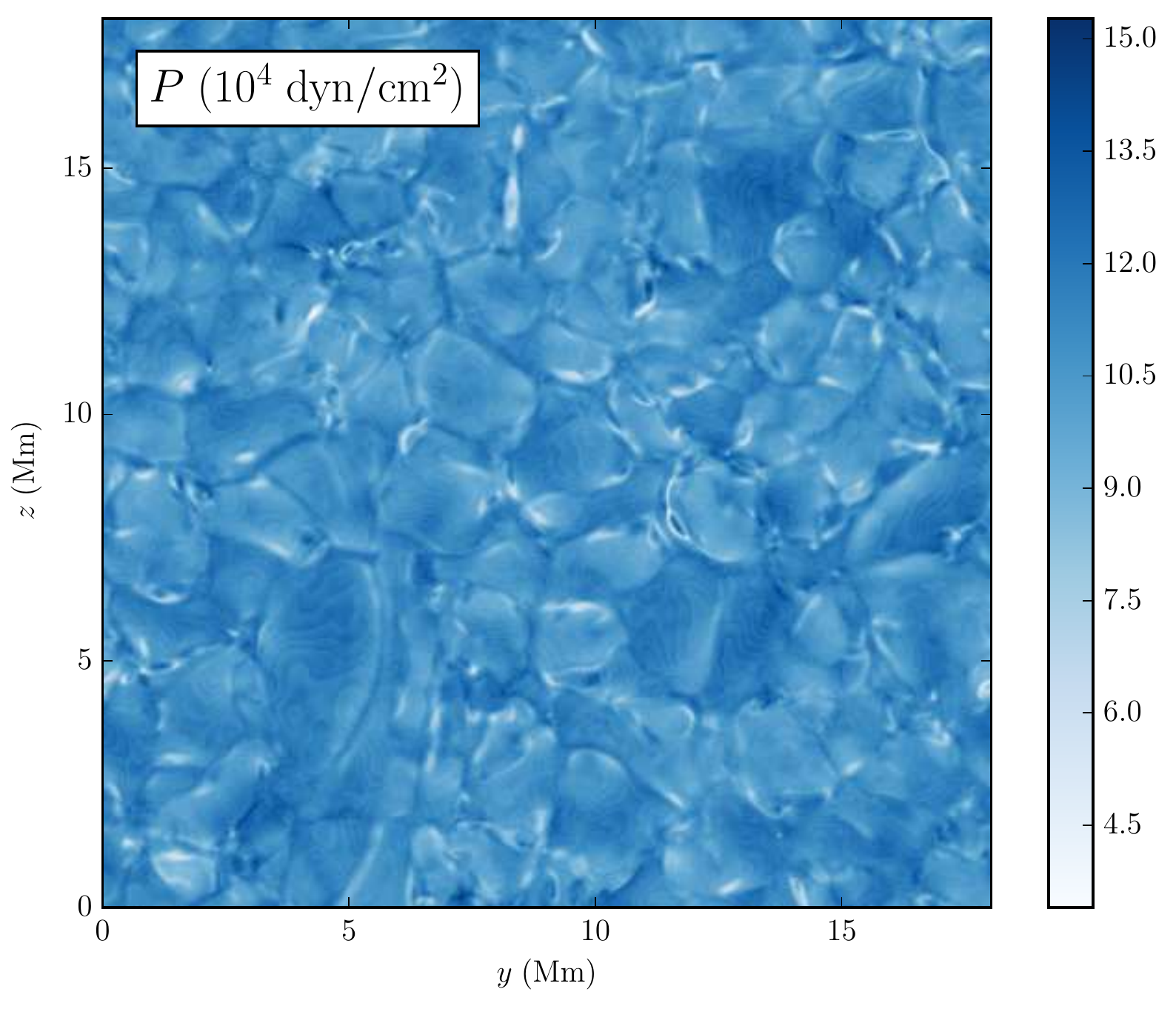}
\includegraphics[width=0.245\textwidth]{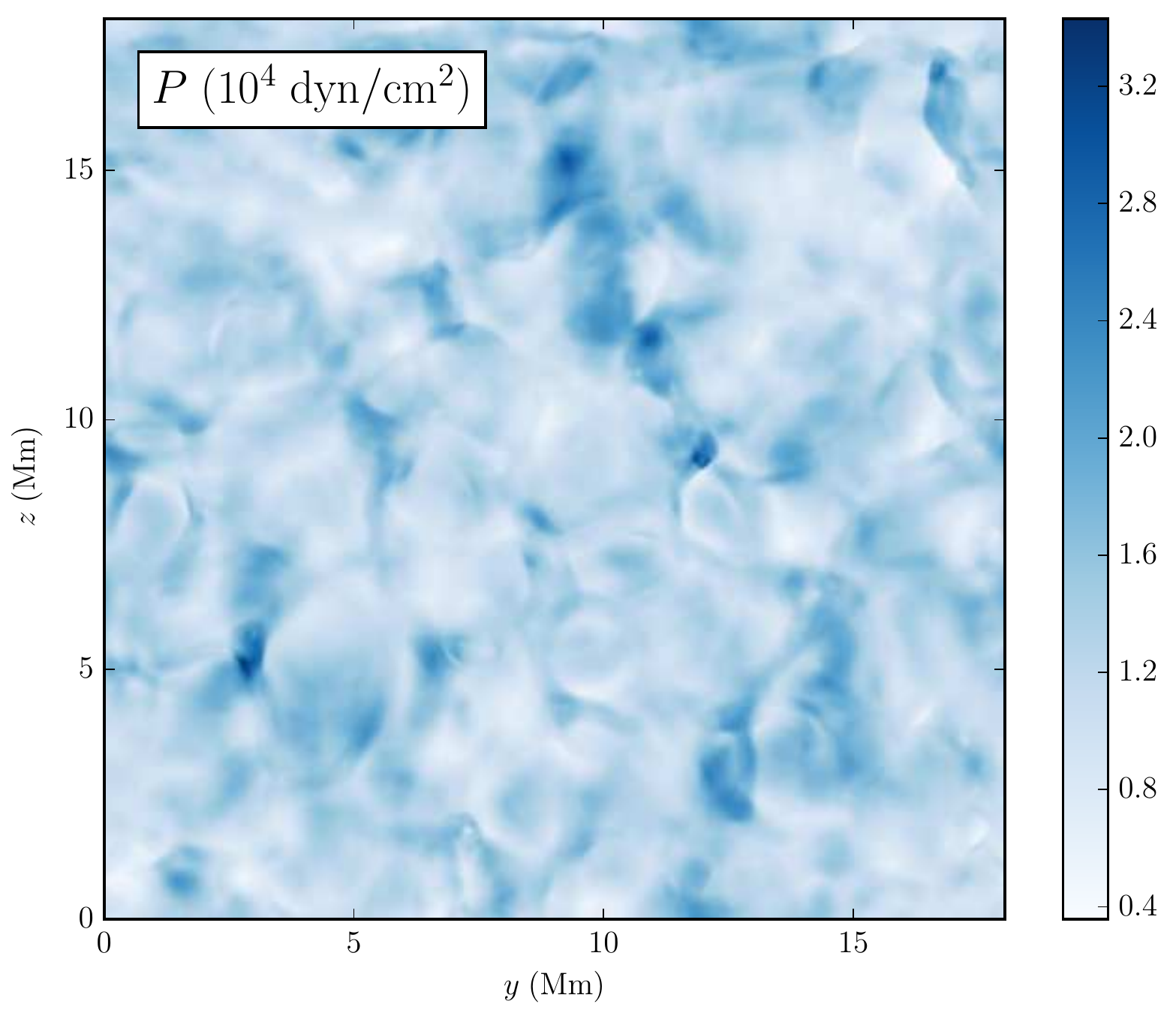}
\caption{Distributions of temperature, intensity, vertical flow velocity, gas density, and gas pressure evaluated at characteristic height levels in the photosphere. The variation of the horizontal temperature distribution $T(y,z)$ at the visible surface is much smaller than the one at the zero-height level as hot granules are observed higher up due the temperature-increasing $\mbox{H}^-$ opacity while cooler gas in the intergranular lanes is observed at considerably deeper layers \hfill\break}
\label{fig:height_dependent_images_of_TurhoP}
\end{figure*}

\begin{table*}
\caption{Mean and extreme values of various quantities at the solar surface for a given time step. Each quantity is given separately for granular and intergranular regions on a horizontal plane corresponding to horizontally averaged optical depth unity and on the optical depth unity isosurface evaluated from Eq.~(\ref{eqn:tau1}).}
\begin{centering}
\begin{tabular}{lrrrr}
\hline \hline  
 & Upflow & Upflow& Downflow & Downflow\tabularnewline
Quantity  & const $\langle x|_{\tau=1} \rangle_{y,z}$ & $\tau=1$ isosurface & const $\langle x|_{\tau=1} \rangle_{y,z}$ & $\tau=1$ isosurface\tabularnewline
\hline
$A_\mathrm{rel}$ & 0.6142325 & 0.6225696 & 0.3857675 & 0.3774304\tabularnewline
$T_\mathrm{mean}\ \mathrm{(K)}$ & 6774.064 & 6489.807 & 5846.059 & 6026.084\tabularnewline
$T_\mathrm{max}\ \mathrm{(K)}$ & 10\,498.42 & 7315.655 & 7893.692 & 6915.974\tabularnewline
$T_\mathrm{min}\ \mathrm{(K)}$ & 4778.876 & 5416.145 & 4904.417 & 5343.949\tabularnewline
$\varrho_\mathrm{mean}\ \mathrm{(10^{-7}\ g/cm^3)}$ & 2.360992 & 2.517944 & 2.460644 & 2.763120\tabularnewline
$\varrho_\mathrm{max}\ \mathrm{(10^{-7}\ g/cm^3)}$ & 4.041061 & 3.956815 & 4.656799 & 3.881417\tabularnewline
$\varrho_\mathrm{min}\ \mathrm{(10^{-8}\ g/cm^3)}$ & 8.249153 & 7.930662 & 6.471923 & 8.340405\tabularnewline
$I_\mathrm{mean}\ \mathrm{(erg\,cm^{-2}\,s^{-1})}$ & $5.915369 \times 10^{10}$ & $2.517944 \times 10^{-7}$ & $2.937458 \times 10^{10}$ & $2.763120 \times 10^{-7}$\tabularnewline
$I_\mathrm{max}\ \mathrm{(erg\,cm^{-2}\,s^{-1})}$ & $2.258222 \times 10^{11}$ & $3.956815 \times 10^{-7}$ & $8.080531 \times 10^{10}$ & $3.881417 \times 10^{-7}$\tabularnewline
$I_\mathrm{min}\ \mathrm{(erg\,cm^{-2}\,s^{-1})}$ & $1.842588 \times 10^{10}$ & $7.930662 \times 10^{-8}$ & $1.922034 \times 10^{10}$ & $8.340405 \times 10^{-8}$\tabularnewline
$u_\mathrm{mean}\ \mathrm{(km/s)}$ & $-2.045552$ & $-1.927267$ & 1.978299 & 2.134066\tabularnewline
$u_\mathrm{max}\ \mathrm{(km/s)}$ & $-10.18814$ & $-7.807356$ & 8.128013 & 8.805439\tabularnewline
$u_\mathrm{min}\ \mathrm{(km/s)}$ & $-6.289904 \times 10^{-5}$ & $-5.699339 \times 10^{-5}$ & $1.119909 \times 10^{-5}$ & $6.379750 \times 10^{-5}$\tabularnewline \hline
\end{tabular}
\par\end{centering}
\label{tab:quantities_over_gran_intergran}
\end{table*}

\begin{figure}[htbp]
\centering
% Change png to eps finally but not while still writing the paper. Compiling with original eps scatter images takes incredibly long!
\includegraphics[width=\columnwidth]{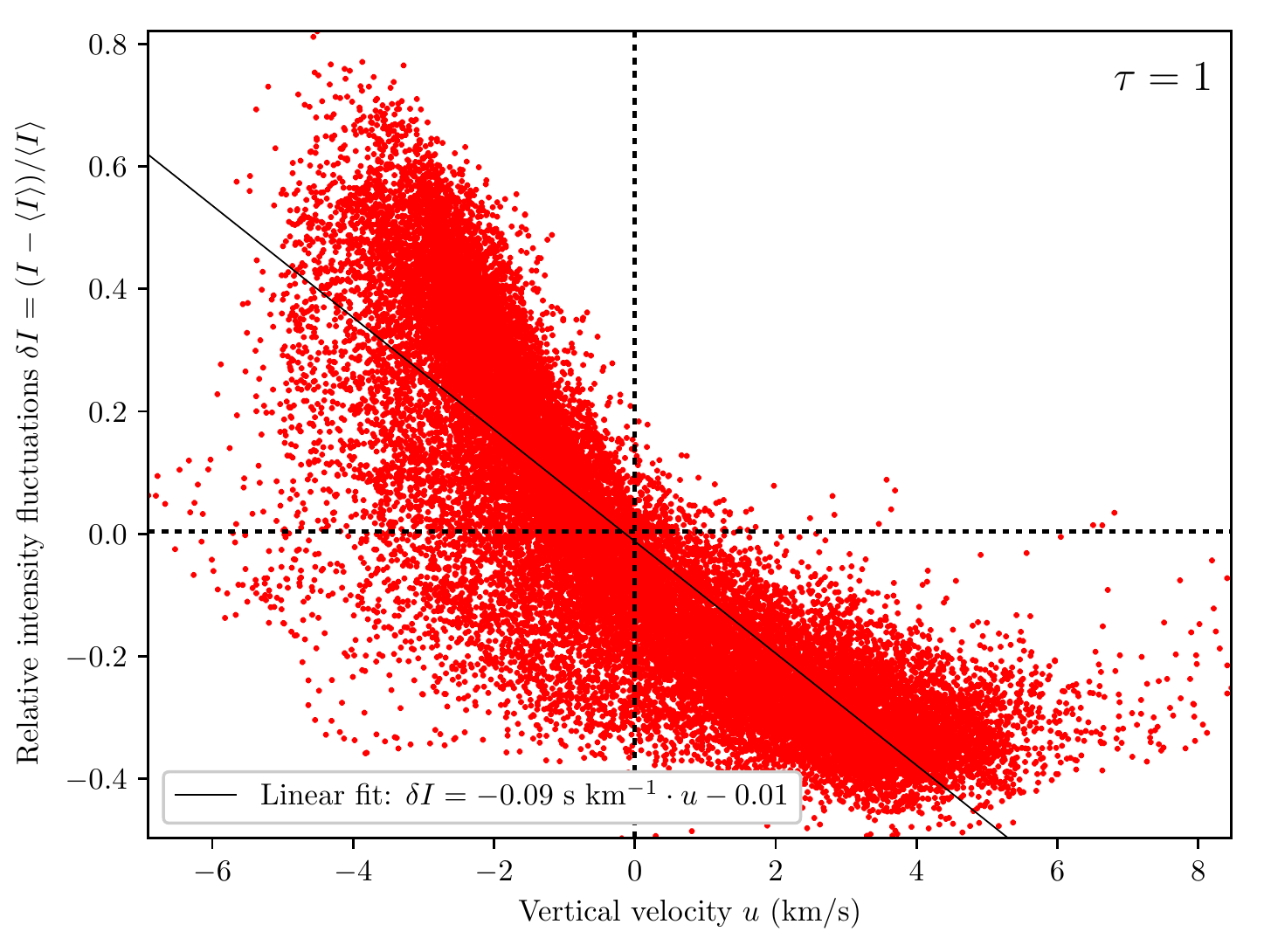}
%\columnbreak
%\end{figure}
%\begin{figure}[H]
\includegraphics[width=\columnwidth]{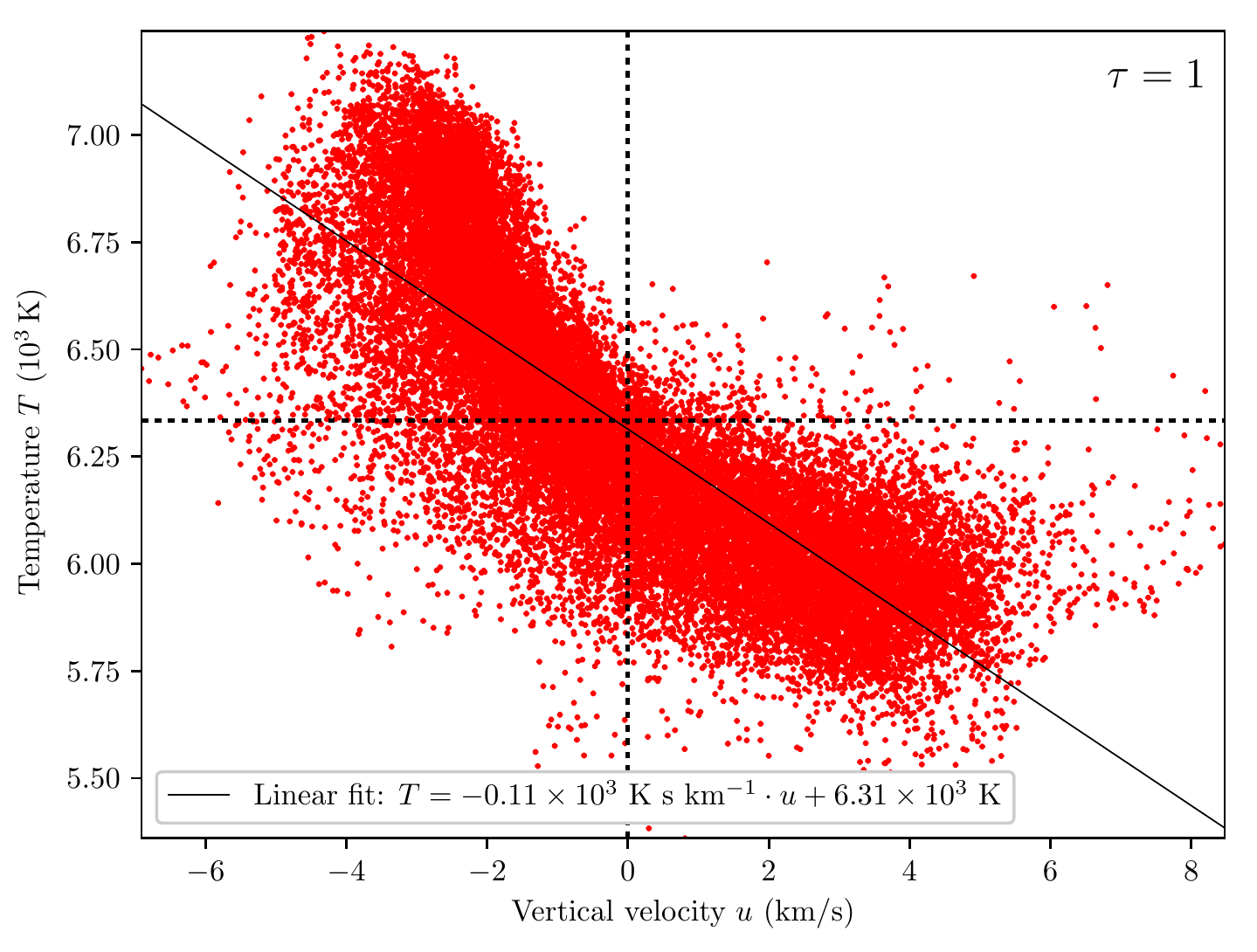}
%\end{figure}
%\begin{figure}[H]
\includegraphics[width=\columnwidth]{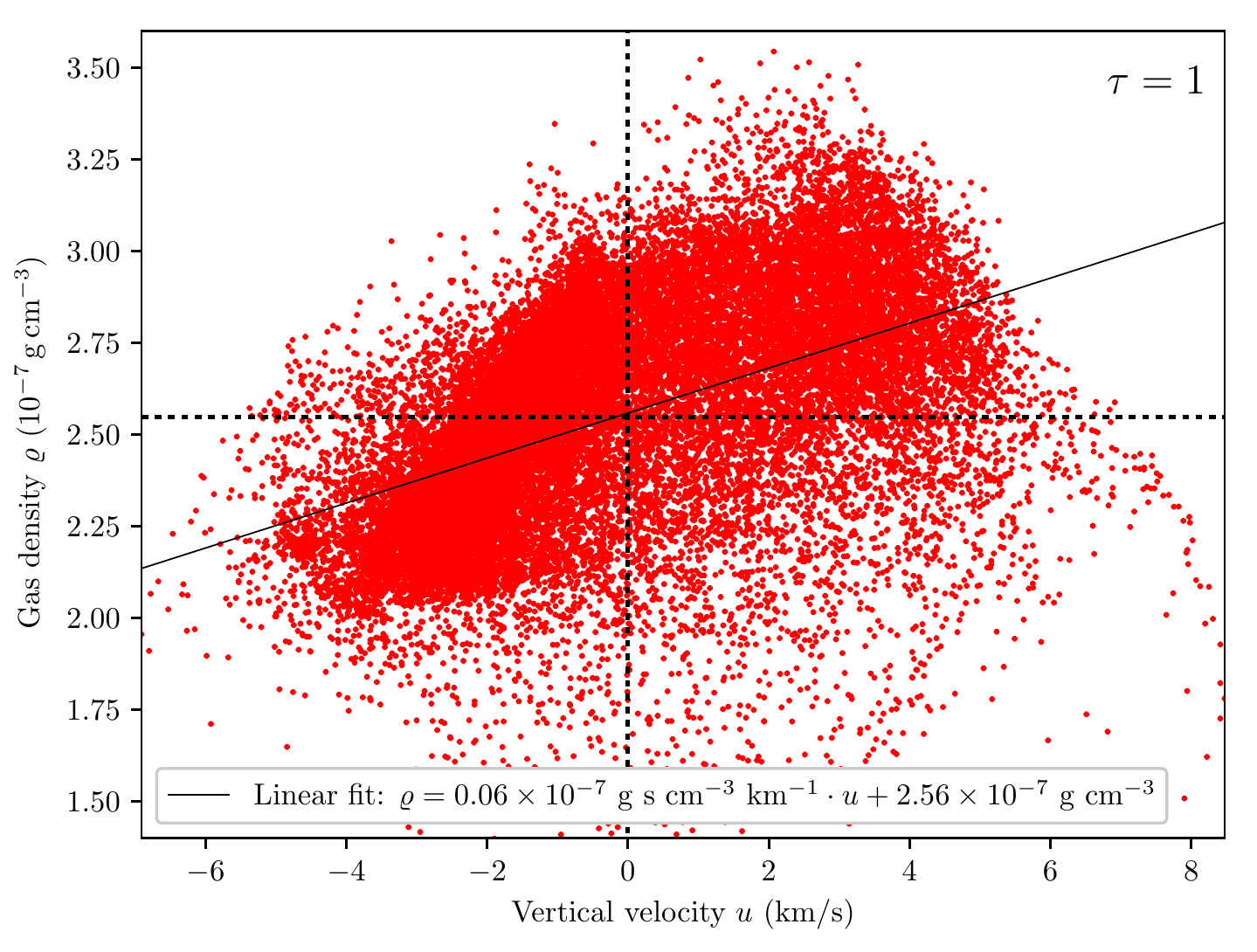}
\caption{Correlation of intensity, temperature, and gas density with the radial flow velocity at the visible surface $\tau=1$}
\label{fig:correlations_at_tau1_surf}
\end{figure}

\section{Results}
\subsection{Stratification of the photosphere}\label{sec:stratification}
Figure~\ref{fig:height_dependent_images_of_TurhoP} shows the height variation of typical model quantities' horizontal distributions observed at a given snapshot in the photosphere. As we can see from the temperature and intensity images in the first and second row, respectively, the temperature maxima at the surface and below are located above the granules, while the gas in the intergranular lanes is some thousand degrees cooler. Higher up in the photosphere---the chosen level of 275~km in the rightmost column of the figure corresponds to a maximum temperature difference in up- and downflows in the middle photosphere as will be discussed in the following sections---the temperatures are inverted due to buoyancy breaking such that the gas above granules is significantly cooler than the coalescing downdrafts in the intergranular lanes. In contrast, the observed intensity in the middle photosphere does not mirror its image at the surface: granules are brighter throughout the photosphere. The intensity distributions peak at the borders of the granular upflows and in particular in regions bordering merging intergranular lanes. As discussed above, temperatures and intensities vary significantly stronger at a given geometric depth than at a surface of constant optical depth. Comparing the intensity and velocity distributions (third row) shows that the highest intensity regions within granules correspond also to the maximum upward directed
flow velocities. Downdrafts are observed to be highest where at least two intergranular lanes meet and the cooled flows of several granules converge. Again, the range of values is here slightly higher at surface level-constant geometric depth, but maximum values for downflow velocities are found at constant optical depth unity, where downflow regions are located in considerably deeper layers. The density and pressure distributions are shown in rows 4 and 5. Below the surface coalescing downflows are significantly denser than the granular upflows and are associated with a lower pressure. While pressure and density images are mirrored in subphotospheric layers, relative density- $\delta\varrho$ and pressure fluctuations $\delta P$ become of comparable size above the transition layer once radiative equilibrium is established. Here, the energy exchange of the perturbations with the surroundings can be considered
isothermal. Above the transition layer we observe a gradual breakdown of the columnar structure. The mean and extreme values of temperature, intensity, gas density, intensity and vertical flow velocity are summed up in Table~\ref{tab:quantities_over_gran_intergran}, each for granular and intergranular regions on constant geometric depth and on the $\tau$-unity isosurface, respectively.

In the following we will put this qualitative description of the spatial variation of model quantities with the granular brightness field as well as with height on a more quantitative basis in terms of a correlation analysis. We start our analysis of the model photosphere by reproducing the most obvious correlation between the vertical flow velocity and the brightness at the solar surface that was found in observations already in the mid-twentieth century \citep[e.g.][]{Stuart1954}. From the leftmost panel of Fig.~\ref{fig:correlations_at_tau1_surf} it is apparent that bright granular areas at the $\tau$-unity isosurface are associated with a negative, i.e. upward vertical fluid flow velocity and vice versa for the darker gas in the intergranular lanes. The correlations of the vertical velocity with temperature (middle panel) and gas density (right panel) show that at the surface the upflows are significantly hotter but less dense compared to the intergranular downdrafts. These trivial dependencies already reveal that the correlation of the vertical flow with the density is weaker than its correlation with the brightness, just as that the allocation of low density gas with upflows is a good one but not unflawed. This incident was already apparent from column 3 of Fig.~\ref{fig:height_dependent_images_of_TurhoP}. The wide dispersion shows that low gas densities are to be found in downflows as well and the strongest downdrafts with $u \approx8~\mbox{km}/\mbox{s}$ even exclusively exhibit low gas densities in contrast with the overall trend.

\subsection{Relative fluctuations of thermodynamic variables in two-component representation}\label{sec:rel_fluctuations}
\begin{figure}
\centering
\includegraphics[width=.98\columnwidth]{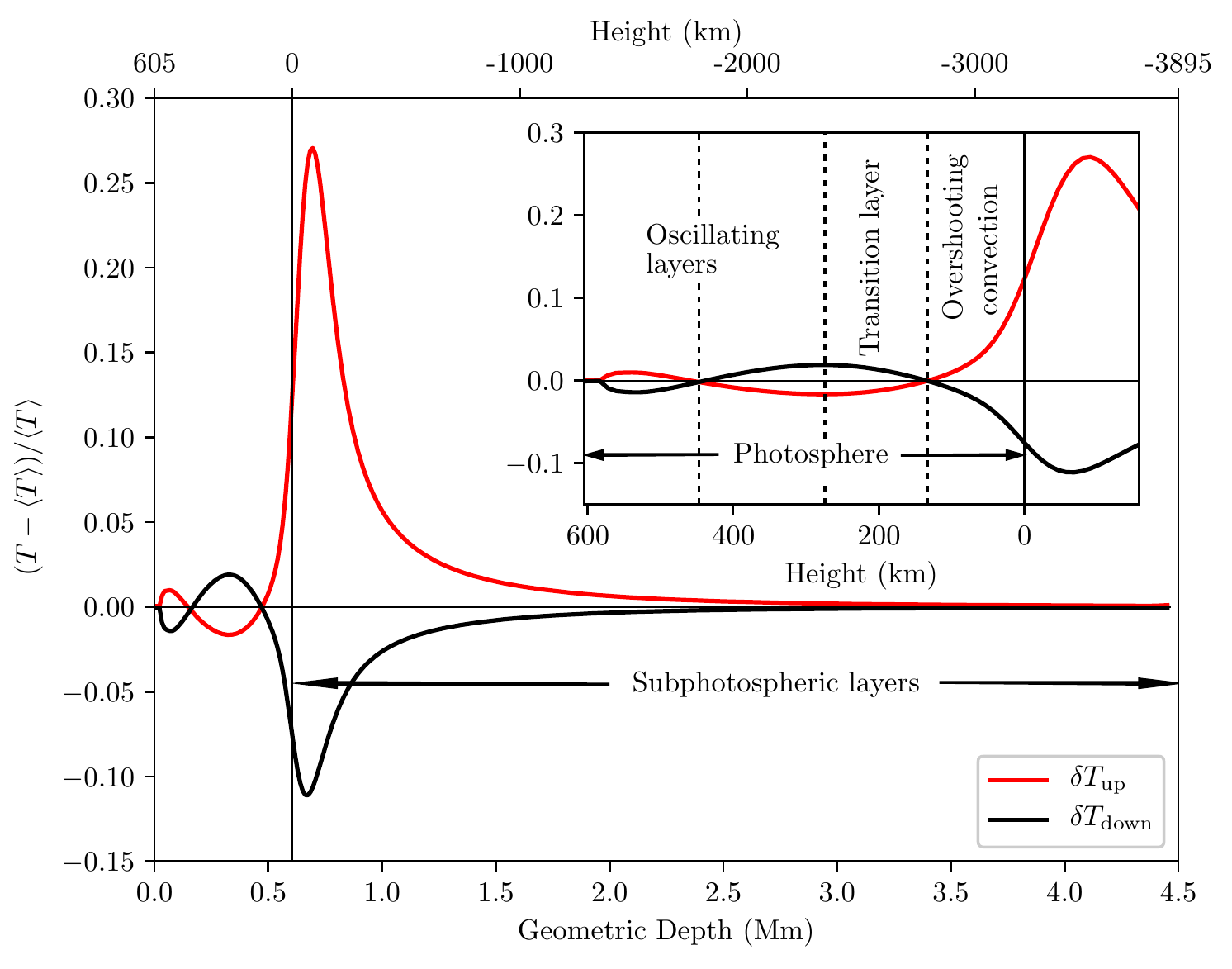}
\caption{Relative temperature fluctuations averaged over up- (\emph{red}) and downflows (\emph{black}) as well as time as a function of height from subphotospheric layers to the convectively stable adjacent photosphere. The \emph{inset} is a zoom of the photospheric region showing the subdivision into layers of convective overshoot above the surface, a transition layer, and oscillating layers in the upper photosphere}\label{fig:temperature_fluctations}
\end{figure}

Relative fluctuations $\delta Q$ of some quantity $Q$ were evaluated according to the procedure described in Sect.~\ref{sec:data_analysis}. The convectively unstable subphotosheric layers are characterized by divergent temperature fluctuations in up- and downflows that are most prominent only 90~km and 70~km, respectively below the photosphere, Fig.~\ref
{fig:temperature_fluctations}. As one moves further upwards into the photosphere, temperature fluctuations in up- and downflows reverse their sign at a height of $\approx 130~\mbox{km}$. This inversion that peaks at $\approx 275~\mbox{km}$ can be attributed to the buoyancy breaking effect explaining the inversion via two mechanisms: First, as the upflowing matter overshoots into the convectively stable region it is bound to lose energy via radiation due to a strong decrease in opacity. Second, as downflows are coalescing into the intergranular lanes, they become compressed and heated. Going still further upwards, the temperature of upflows increases again as these upper layers become compressed by oscillations.
These height levels were found to vary depending on the employed photospheric model: The first reversal of temperature for instance is located in a height range from 100~km based on the
mechanical-radiative energy balance model of the solar granulation by \cite{Musman1976} up to 170~km as observed from the 2-D RHD model of \cite{Gadun1999}. A slight variation is found also in relative height levels such as the distance between the two temperature reversal points that in the latter 2-D model falls short of the distance observed from our modeling by 20\%.

\begin{figure}[htbp]
\centering
\includegraphics[width=0.98\columnwidth]{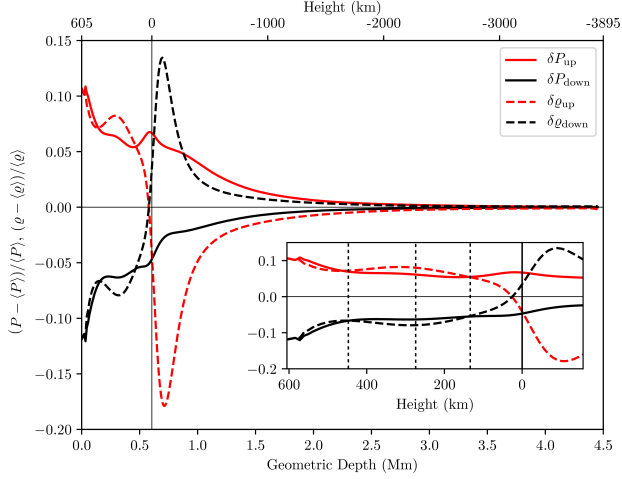}
\caption{Relative fluctuations of pressure (\emph{solid}) and density (\emph{dashed}) averaged over upflows (\emph{red}), downflows (\emph{black}) and time}
\label{fig:pressure_fluctuations}
\end{figure}

Figure~\ref{fig:pressure_fluctuations} shows that absolute pressure fluctuations are significantly greater than absolute density fluctuations $|\delta P| > |\delta\varrho|$ in the lower photosphere and become of roughly equal size above the first reversal of temperature fluctuations $|\delta P| \approx|\delta \varrho|$, indicating radiative equilibrium above that level, while directly at the reversal points their concurrence is exact, $|\delta P|=|\delta\varrho|$.

\begin{figure}[htbp]
\centering
\includegraphics[width=\columnwidth]{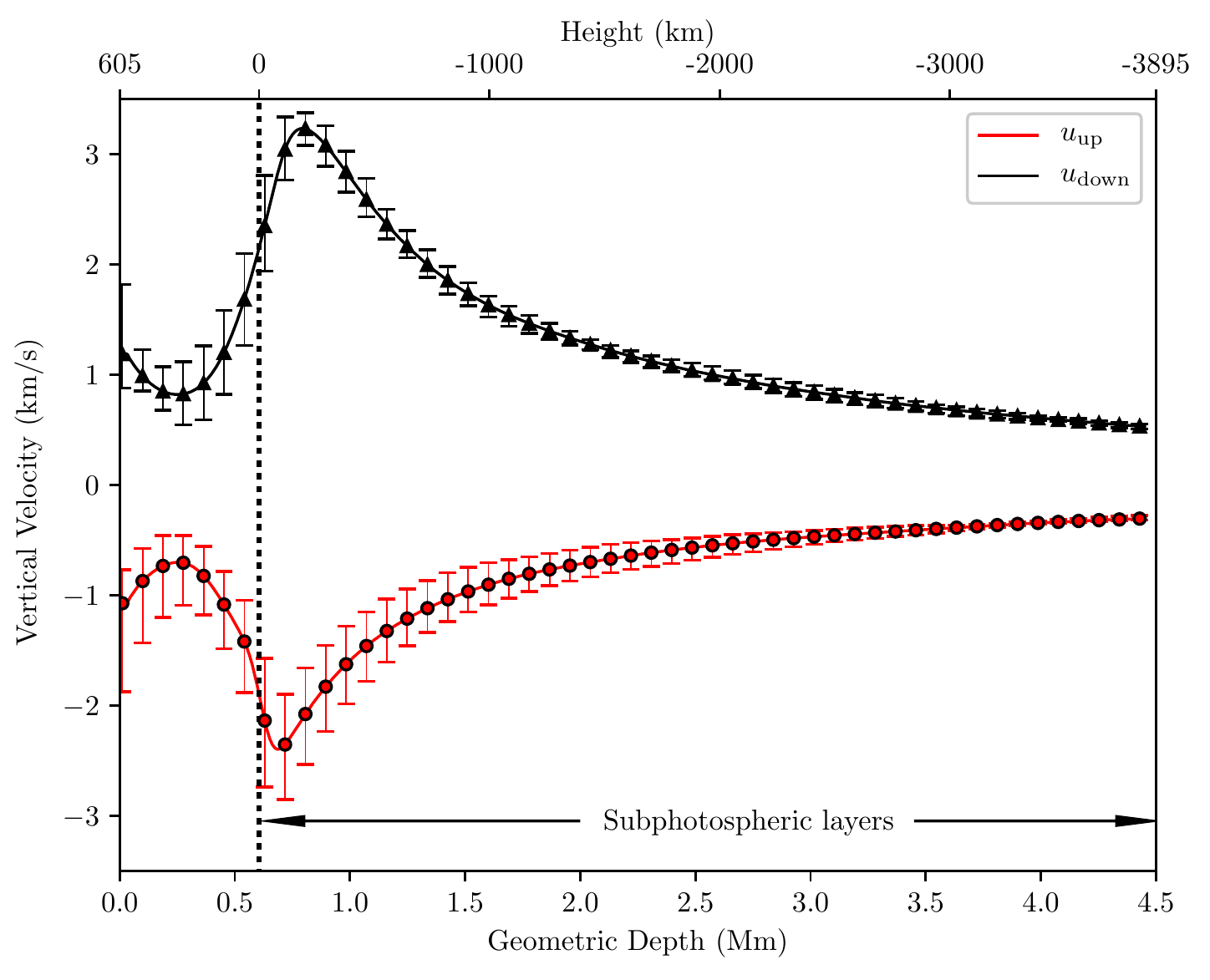}
\caption{Absolute values of the vertical velocity in up- and downflows for the full computational domain.}
\label{fig:velocities}
\end{figure}

Temporally and horizontally averaged vertical velocities evaluated separately for granular and intergranular flows are shown in Fig.~\ref{fig:velocities}. Error-bars indicate the temporal variability of the horizontal mean of the quantity in consideration. The vertical velocity field does not invert across the photosphere.
Maxima of the vertical flow velocity in up- and downdrafts as listed in Table~\ref{tab:quantities_over_gran_intergran} are averaged out and hence are not reflected in this graph. Downdrafts on average have higher velocities throughout the photosphere
as well as in subphotospheric layers. Both, up- and downdrafts reach their maximum flow velocity just below the photosphere, where also the temperature separation of the two flows is most pronounced. In absolute values the average flow velocities of
up- and downdrafts decrease from the surface across the lower photosphere toward the top of the transition zone and rise again in the higher oscillatory layer. As was pointed out
already by \citet{Schwarzschild1948}, acoustic waves above granules proceed essentially without dissipation into the higher photosphere and the perturbations do not affect the temperature there, such that the energy flux in acoustic waves $\varrho \boldsymbol{u}^2 c_\mathrm{s}$ with sound speed $c_\mathrm{s}$ can be considered constant, explaining the rise of the average flow velocity with decreasing density in the oscillatory layers. This requires flow velocities well below a critical value $u^2 < u_\mathrm{crit}^2 = \beta P/\varrho$ with some fraction $\beta$ of the order of 0.1. While by and large the same argument also applies to the downflows, one should not leave unmentioned that locally fast downdrafts exceed the critical flow speed and even shocks are found sparsely scattered in the intergranular lanes.

\subsection{Local and two-point model quantity correlations}\label{sec:local_and_2point_correlations}
We first discuss local or one-point linear correlations between various model quantities before turning to (relative) two-point correlations for which one quantity is fixed at the solar surface while the second one is varied with height.

\begin{figure}[htbp]
\centering
\includegraphics[width=\columnwidth]{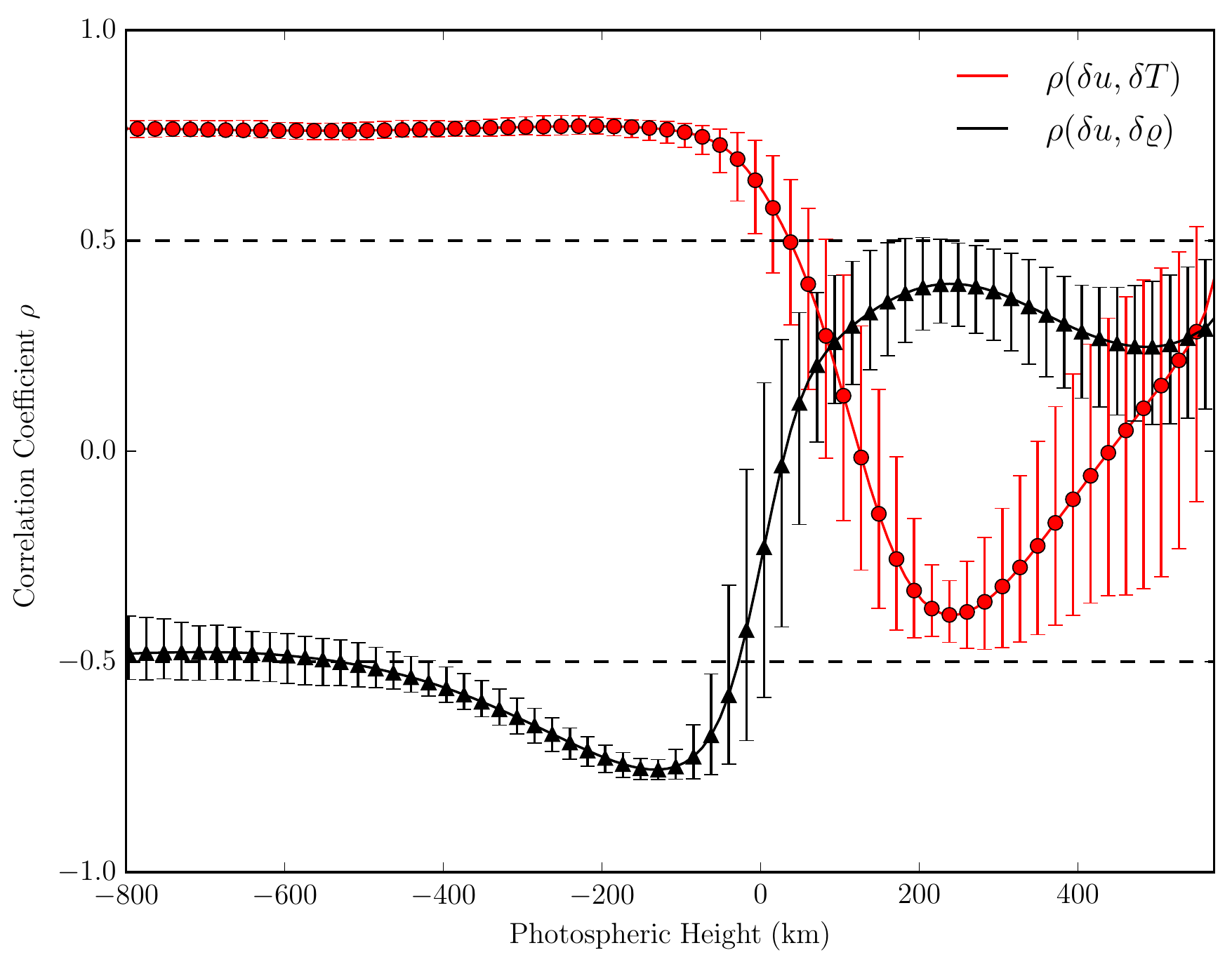}
\includegraphics[width=\columnwidth]{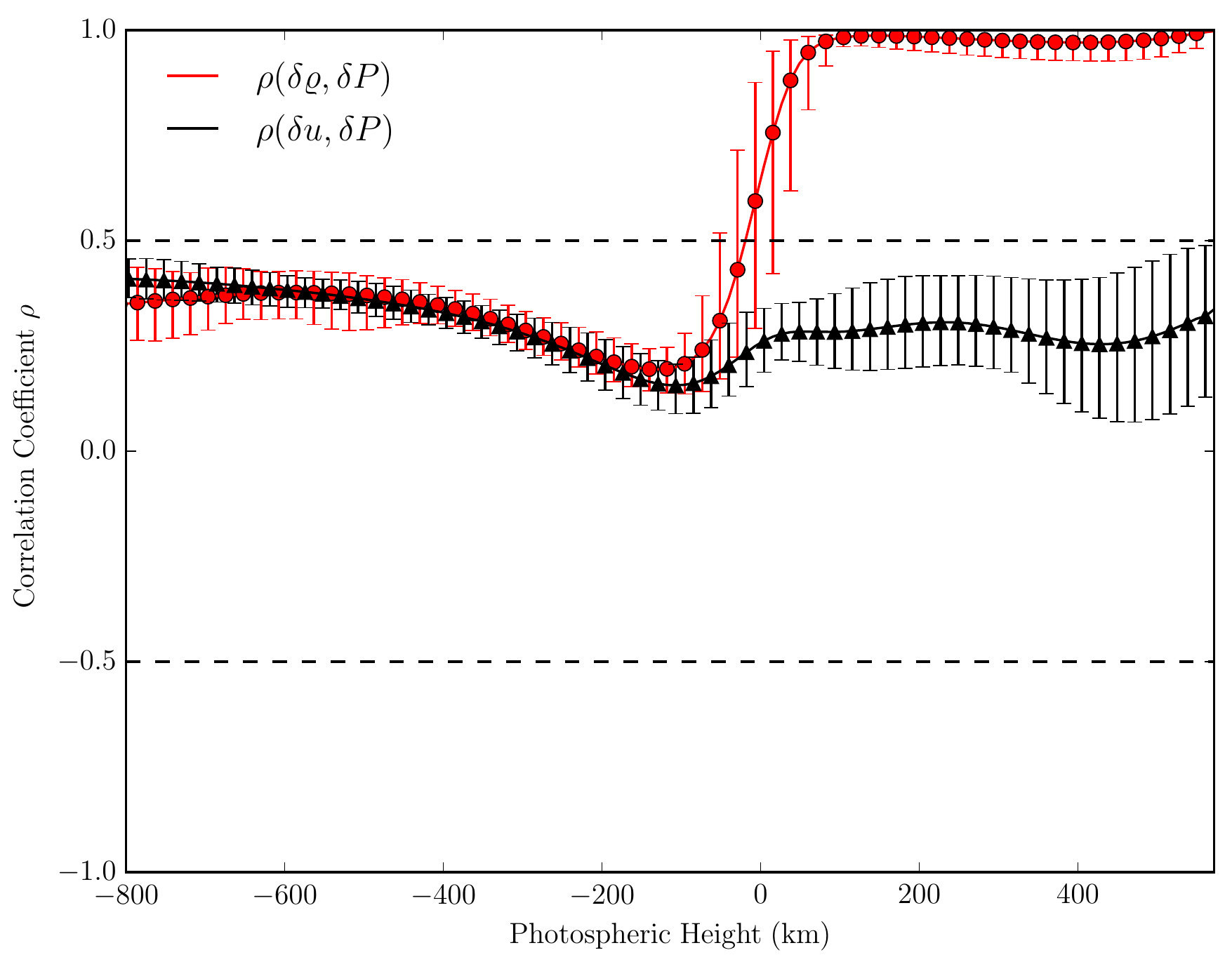}
\caption{One-point correlations (averaged over time) as a function of height in the photosphere. Error-bars indicate the temporal variation of these correlation functions. \textit{Top:} Correlation of fluctuations of the vertical velocity with temperature (\emph{red}) and gas density (\emph{black}). \textit{Bottom:} Correlation of fluctuations of the gas density with gas pressure (\emph{red}) and the vertical velocity with gas pressure}\label{fig:1p_corr}
\end{figure}

The one-point correlation between the vertical velocity- and temperature fluctuations $\rho(\delta u, \delta T)$ pictured in the upper panel of Fig.~\ref{fig:1p_corr} again reflects the temperature reversals discussed before. As expected, a strong correlation is found in the subphotospheric convective layers and an anticorrelation in the most part of the photosphere due to the overcooling of the photospheric matter in optically thin layers. We also observe from this figure that the anticorrelation of the vertical velocity- and gas density fluctuations $\rho(\delta u,\delta\varrho)$, characteristic for convectively unstable regions, reverses its sign at a height of $\approx 35~\mbox{km}$ up from where a pursuing positive correlation is found. The gas density and pressure are strongly correlated throughout the photosphere, see the lower panel of Fig.~\ref{fig:1p_corr}.

\begin{figure}[htb]
\centering
\includegraphics[width=\columnwidth]{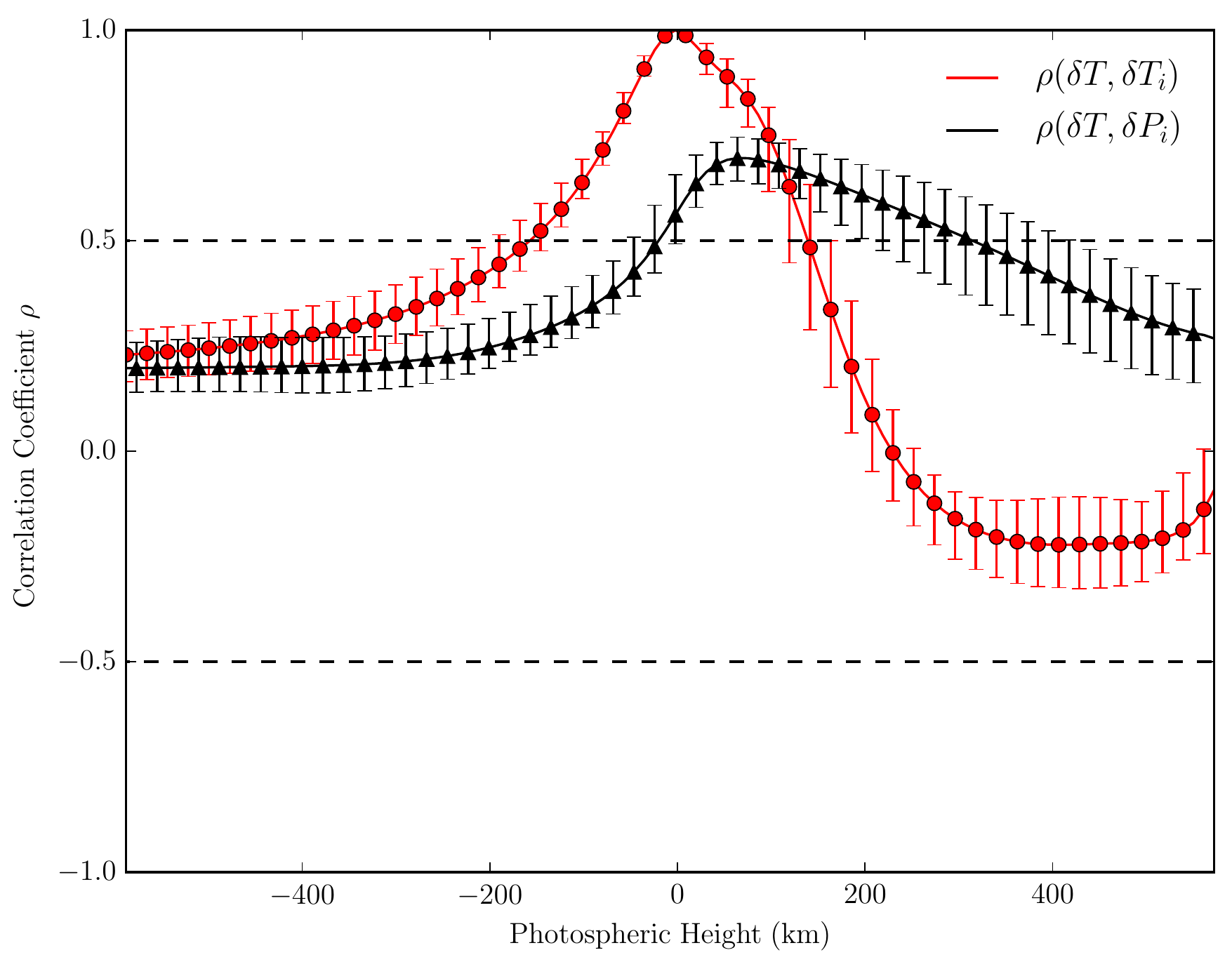}
\includegraphics[width=\columnwidth]{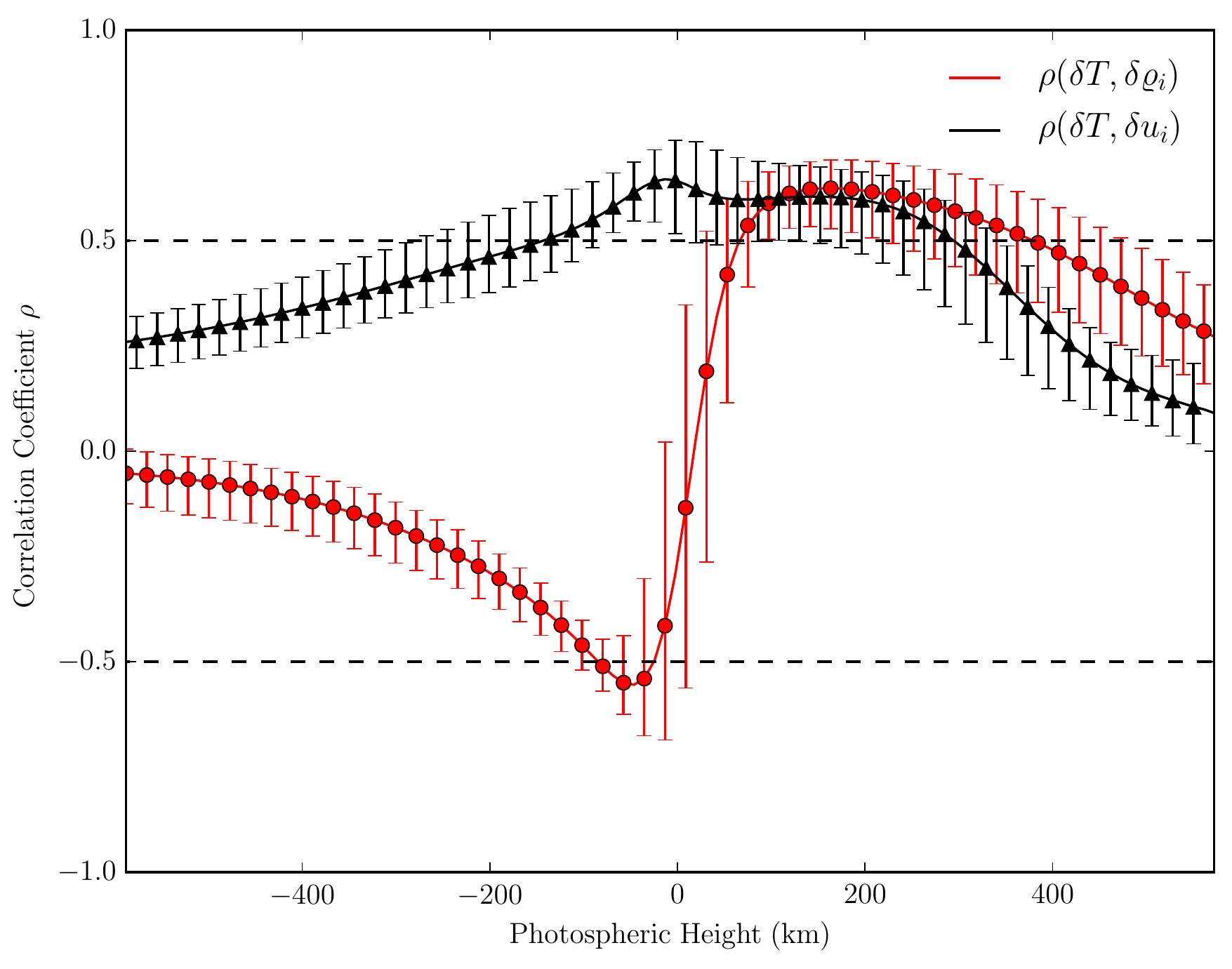}
\caption{\textit{Top:} Two-point correlations of temperature fluctuations (fixed at $x|_{\tau=1}$) and running temperature and pressure fluctuations, respectively. \textit{Bottom:} Same for temperature fluctuations and running density as well as vertical velocity fluctuations}
\label{fig:2p_corr}
\end{figure}

A high correlation of temperature fluctuations at the solar surface with temperature fluctuations at varying heights, $\rho(\delta T, \delta T_i)$ is found from subphotospheric layers to the lower photosphere as shown in the upper panel of Fig.~\ref{fig:2p_corr}. As~expected, the correlation steeply drops from the solar surface (where it is of course exact) upwards until it becomes negative below the temperature reversal of up- and downflows. Throughout the region of temperature reversal the anticorrelation is strongest and is found to decrease again above the second temperature reversal point yet without turning positive again. In subphotospheric layers the hotter and brighter gas of the granulation cells is less dense as can also be seen by comparing the temperature and density images at $h=-100~\mbox{km}$ in Fig.~\ref{fig:height_dependent_images_of_TurhoP}, why not surprisingly a strong anticorrelation $\rho(\delta T, \delta \varrho_i) \approx -0.6$ is found where the temperature separation of up- and downflows has reached its maximum level few ten kilometers below the surface, see lower panel of Fig.~\ref{fig:2p_corr}. As the temperature reverses from here, the correlation function becomes positive very fast, reaching a high level of $\rho(\delta T, \delta \varrho_i) \approx 0.6$ in the transition region between thermal convection and the oscillating layers. Here, hotter and brighter gas is found in the intergranular lanes, cf. also the temperature image at $h = 275~\mbox{km}$ in Fig.~\ref{fig:height_dependent_images_of_TurhoP} which almost mirrors the temperature distribution at $h = -100~\mbox{km}$. As the temperature separation in up- and downflows diminishes above that layer again, so also decreases the correlation in the above regions where the columnar structure is no longer present.

The gas pressure is higher in granular upflows throughout subphotospheric and photospheric regions resulting in an entirely positive correlation $\rho(\delta T, \delta P_i) \geq 0.2$ (black curve in the leftmost panel of Fig.~\ref{fig:2p_corr}). Of course, throughout and above the transition layer, where due to a radiative equilibrium $\delta P \approx \delta \varrho$, the correlation functions of temperature fluctuations with $\delta P_i$ and $\delta \varrho_i$, respectively show a similar run with height.

From the images of the vertical velocity in Fig.~\ref{fig:height_dependent_images_of_TurhoP} it is apparent, that $u$ is always negative above granules (corresponding to upflows) and positive in the intergranular lanes. In the oscillatory layers however that clear allocation becomes increasingly blurred as the columnar structure breaks down. Quantitatively this is shown by the correlation function $\rho(\delta T, \delta u_i)$ (black curve in the lower panel of Fig.~\ref{fig:2p_corr}). The high correlation in subphotospheric layers and the lower photosphere drops notably to insignificant values above the transition layer.

Finally, we discuss the two-point correlation of the intensity fluctuations at the surface with the mean opacity fluctuations at varying heights, $\rho(\delta I, \delta\kappa_i)$, cf. the red curve in Fig.~\ref{fig:2p_corr_no2}. It is striking that up to the surface this correlation function almost coincides with the correlation of intensity fluctuations with temperature fluctuations, which, as was already argued by \cite{Gadun2000}, is due to the ionization of hydrogen which is strongly temperature dependent as $\mbox{H}^-$ ions primarily cause the absorption of photons here. In higher layers where due to the still high temperatures basically ionized metals are mainly accountable for the absorption of photons, the fluctuations of the opacity are no longer sensitive to temperature fluctuations. The authors' claim based on their 2-D RHD model that up from here $\rho(\delta I, \delta\kappa_i)$ is closely following the correlation function $\rho(\delta I, \delta P_i)$ could not be reproduced with our recent model, Fig.~\ref{fig:2p_corr_no2}.

\begin{figure}[h]
\centering
\includegraphics[width=\columnwidth]{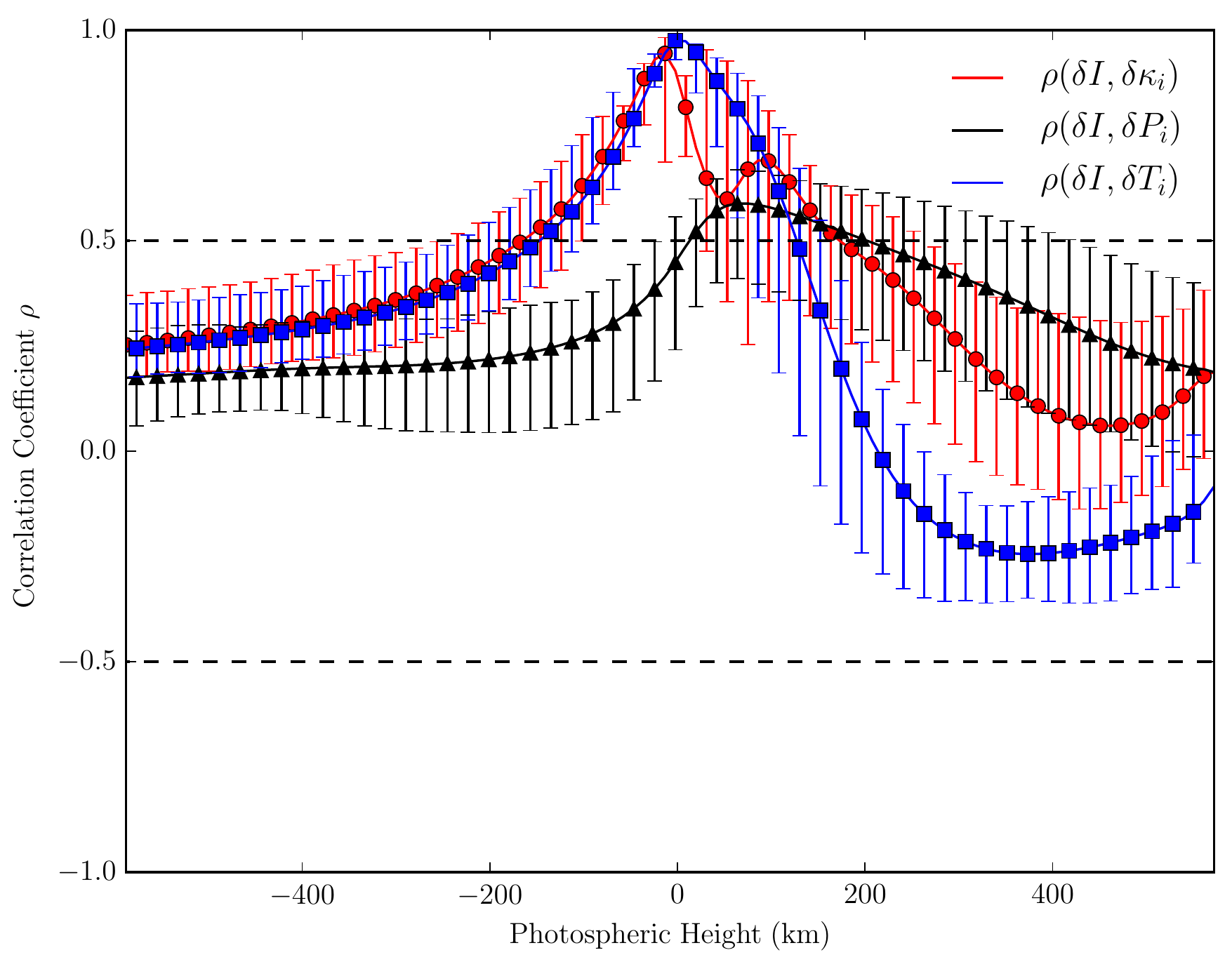}
\caption{Correlation of intensity fluctuations (fixed at $x|_{\tau=1}$) with running opacity-, pressure- and temperature fluctuations}
\label{fig:2p_corr_no2}
\label{fig:2p_corr_no2}
\end{figure}

Altogether these data provide the necessary information for outlining the overall vertical structure of the photosphere (see also Table~\ref{tab:structure}): Due to the rapid decrease in opacity, see right panel of Fig.~\ref{fig:tau_kappa_run}, radiative cooling quickly gains in importance as one proceeds from subphotospheric layers across the $\tau=1$ isosurface. Following the argument of \citet{Gadun2000}, we can interpret the height level where $\rho(\delta u,\delta\varrho)$ turns positive as the top of the thermal convection zone which thus extends to some ten kilometers above the solar surface. While the adjacent layer still exhibits positive and negative temperature fluctuations in up- and downflows, the further increasing positive correlation $\rho(\delta u, \delta\varrho)$ indicates that this layer is no longer convectively unstable although thermal convective upflows overshoot into this region up to a height of $\approx
130~\mbox{km}$, where a radiative equilibrium is established and temperature fluctuations reverse their sign. The columnar structure of thermal convection persists up to a height of $\approx275~\mbox{km}$ where the temperature inversion is most pronounced. The adjacent upper photosphere is governed by acoustic oscillations and the columnar structure is no longer observable. Comparing these findings with the ones of pioneering RHD simulations of the solar granulation developed by \citet{Musman1976} or the 2-D model of \citet{Gadun2000}, we find a good agreement of the qualitative runs of correlation functions; finally we were able to update the structural division levels based on our up-to-date numeric model.

\begin{table}
\begin{centering}
\begin{tabular}{lc}
\hline \hline 
Photospheric layers & Approx. Height (km) \tabularnewline \hline
Max. temperature fluct. & $-90$ (up)/ $-70$ (down) \tabularnewline
Convectively instable layers & $<35$ \tabularnewline
Convective overshoot & up to $130$\tabularnewline
Transition layer & $130 \leq h \leq 275$\tabularnewline
Oscillatory layer & $> 275$\tabularnewline
2$^\text{nd}$ temperature reversal & $450$\tabularnewline
\hline
\end{tabular}
\par\end{centering}
\caption{Photospheric stratification classified by its dynamically distinguished layers.}
\label{tab:structure}
\end{table}

\section{Discussion}\label{sec:discussion}
We introduce the radiation hydrodynamics code \texttt{ANTARES} that we applied to the study of the solar granulation and that has not yet received much attention in the Solar Physics community. We used correlation analysis to examine the vertical stratification of the photosphere and determined height levels subdividing the photosphere in layers that exhibit characteristic dynamics of their own: The subphotospheric layers up to a height of $\approx35~\mbox{km}$ above the solar surface were found to be convectively
unstable. Convective upflows overshoot further into the lower photosphere into a height of $\approx 130~\mbox{km}$, where temperature fluctuations in up- and downflows coincide and become exactly zero. The overlying layer is a transition region between the convective and oscillatory regimes. Within its roughly 145~km extension the horizontal distribution of the gas temperature mirrors the one at the photospheric footpoints and below. That inversion peaks at the top of this layer. Further up the columnar structure of the photosphere gradually breaks down as one proceeds into the oscillation-controlled higher photosphere. This rough schematic structure confirms findings from previous models \citep[e.g.][]{Gadun2000,Musman1976} and spectral observations, \citep[e.g.][]{Nesis1988,Karpinsky1990}, while with the present model some further accuracy to the sensitively model-dependent height levels is added.

The WENO scheme implemented in \texttt{ANTARES} avoiding oscillatory solutions at discontinuities which otherwise occur due to the interpolation of discrete data in finite
volume methods is particularly useful for the ongoing study of shocks which are observed in the intergranulum of our model photospheres and for the study of acoustic oscillations in the scope of RHD. We intend to further investigate photospheric wave excitation and propagation by the application of wavelet and wavepacket analysis and to
quantify the associated energy transfer through the photosphere.

As this RHD-code is currently heavily under development with an imminent RMHD upgrade to be released, we intend to soon present further model results and focus on photospheric, small-scale, intergranular rotating plasma jets that have been detected and studied in our RHD simulations \citep{Lemmerer2016} but whose supposed contribution to the chromospheric and coronal heating via the generation of MHD kink waves or torsional Alfv\'{e}n waves relies on testing our assumptions by studying equivalent RMHD model photospheres.

\acknowledgments
This work was supported by the European Commissions FP7 Capacities Program under the Grant Agreement number 312495 and the Austrian Science Fund (FWF) 27765. The model calculation is carried out with support of the VSC project P70068. The work of TVZ was supported by by the Austrian “Fonds zur F\"orderung der wissenschaftlichen Forschung” (FWF) project P 28764-N27 and by the Shota Rustaveli National Science Foundation project DI-2016-17.

\bibliographystyle{abbrv}
\bibliography{SolarPhysics}

\end{document}